\documentclass[%
 reprint,
superscriptaddress,
nofootinbib,
 amsmath,amssymb,
]{revtex4-2}

\usepackage{natbib}[sort&compress]
\usepackage{comment}
\usepackage[english]{babel}
\usepackage[utf8]{inputenc}
\usepackage{graphicx}
\usepackage[colorinlistoftodos, color=green!40, prependcaption]{todonotes}
\usepackage[pdftex, pdftitle={Article}, pdfauthor={Author}]{hyperref} 
\usepackage{tabularx}
\usepackage{subcaption}
\usepackage{ragged2e}
\usepackage{diagbox}
\newcommand{\bra}[1]{\langle #1|}
\newcommand{\ket}[1]{|#1\rangle}

\begin{document}
\title{Dissipation-Assisted Steady-State Entanglement Engineering based on Electron Transfer Models}
\author{Mingjian Zhu}
\email{mz40@rice.edu}
\affiliation{Department of Physics and Astronomy, and Smalley-Curl Institute, Rice University, Houston, TX 77005, USA}
\author{Visal So}
\affiliation{Department of Physics and Astronomy, and Smalley-Curl Institute, Rice University, Houston, TX 77005, USA}
\author{Guido Pagano}
\affiliation{Department of Physics and Astronomy, and Smalley-Curl Institute, Rice University, Houston, TX 77005, USA}
\author{Han Pu}
\email{hpu@rice.edu}
\affiliation{Department of Physics and Astronomy, and Smalley-Curl Institute, Rice University, Houston, TX 77005, USA}
\begin{abstract}
    We propose a series of dissipation-assisted entanglement generation protocols that can be implemented on a trapped-ion quantum simulator. Our approach builds on the single-site molecular electron transfer (ET) model recently realized in the experiment [So {\it et al.} Sci. Adv. {\bf 10}, eads8011 (2024)]. This model leverages spin-dependent boson displacement and dissipation controlled by sympathetic cooling. We show that, when coupled to external degrees of freedom, the ET model can be used as a dissipative quantum control mechanism, enabling the precise tailoring of both spin and boson steady states of a target sub-system. We derive simplified analytical formalisms that offer intuitive insights into the dissipative dynamics. Using realistic interactions in a trapped-ion system, we develop a protocol for generating $N$-qubit and $N$-boson $W$ states. Additionally, we generalize this protocol to realize generic $N$-qubit Dicke states with tunable excitation numbers. Finally, we outline a realistic experimental setup to implement our schemes in the presence of noise sources.
\end{abstract}
\date{\today }
\maketitle
\section{Introduction}
Entanglement is an essential resource for quantum information, underpinning the foundation of numerous quantum gates, algorithms, and protocols for computation and communication \cite{nielsen2001quantum,barenco1995elementary,grover1996fast,bennett2014quantum,RevModPhys.81.865}. Consequently, engineering high-fidelity entangled states becomes a crucial task for quantum information processing platforms. The robustness of an entanglement generation protocol is often deteriorated by dissipation due to the system's interaction with external environments \cite{harrington2022engineered}. However, with a carefully designed system and environment, dissipation can conversely drive the system to a steady state with desired coherence and entanglement, enhancing the robustness of the protocol \cite{kraus2008preparation,verstraete2009quantum}. Such dissipation-driven protocols have been designed and implemented on different platforms, including cavity QED \cite{plenio1999cavity,kastoryano2011dissipative,oliveira2023steady,tokman2023dissipation}, atomic ensembles \cite{PhysRevA.83.052312,roghani2018dissipative,li2020periodically,zhao2024dissipative}, and trapped-ion systems \cite{lin2013dissipative, horn2018quantum,cole2022resource,Malinowski2022,shao2018engineering,huelga2012non,barreiro2011open,Morigi2015,reiter2016scalable}. 

Previous protocols proposed and implemented in trapped-ion systems were based on several commonly used techniques available in the trapped-ion toolbox, such as resolved sideband transitions, optical pumping \cite{leibfried2003singleion, Monroe1995}, two-qubit M{\o}lmer-Sorensen (MS) interactions \cite{sorensen2000entanglement}, and sympathetic cooling \cite{kielpinski2000sympathetic}.
The protocol realized in Ref.~\cite{lin2013dissipative} prepares a singlet Bell state utilizing sideband transition combined with two types of dissipative processes: engineered spontaneous decay channels for resetting electronic state population and sympathetic cooling for removing motional excitations. Ref.~\cite{horn2018quantum} improves this approach using quantum optimal control theory, and the improved version is realized experimentally in Ref.~\cite{cole2022resource}. 
Other schemes Ref.~\cite{shao2018engineering, Malinowski2022} avoid the loss of fidelity due to non-ideal cooling by using controlled spontaneous decay as the only dissipation channel to generate a Bell state. Finally, a scheme proposed in \cite{huelga2012non} can also be used to prepare a singlet state in two qubits by coupling them to a non-Markovian environment composed of two damped local bosonic modes. 

Similarly, it has been shown that dissipation-driven methods can generate many-body entanglement: in Ref.~\cite{barreiro2011open}, a four-qubit Greenberger-Horne-Zeilinger (GHZ) state is dissipatively prepared using a quantum circuit that incorporates a sequence of two-qubit MS gates and an ancillary qubit cyclically reset by optical pumping. The scheme developed in Ref.~\cite{Morigi2015} applies a pulse sequence consisted of nearest neighbor spin-spin MS interaction, sideband transitions, and sympathetic cooling to drive the system into a $N$-spin entangled antiferromagnetic state.  Ref.~\cite{reiter2016scalable} proposes a general scheme to generate many-body GHZ states utilizing sideband transitions and engineered spontaneous decay channels.

In this study, we focus on developing a new set of protocols for preparing entangled states in trapped-ion systems. 
Our proposed scheme is based on a mechanism distinct from those of the previously proposed protocols. Our method draws inspiration from a recent trapped-ion experiment that simulates single-site molecular electron transfer (ET) with controlled dissipation \cite{so2024trapped} and, as we will show later, it is fundamentally connected to the non-Markovian properties of the scheme proposed in \cite{huelga2012non}. ET is essential in many physical and biochemical processes, such as self-exchange reactions and photosynthesis \cite{mohseni2014quantum,marcus1985electron,van2000photosynthetic}. The single-site ET system can be modeled by a spin-1/2 qubit (we refer to it as the ``control qubit" in later parts of the article) coupled to a damped bosonic mode \cite{garg1985effect, wolynes1987dissipation}. Specifically, the spin degree of freedom models a pair of electronic states (donor and acceptor) coupled to a reaction coordinate under friction, encoded as the damped bosonic mode. The dynamical properties of the perturbative regime, which is characterized by a set of nearly discrete resonant transfers \cite{so2024trapped,schlawin2021continuously}, can be particularly useful for steady-state engineering. 

As we demonstrate in this work, if a target quantum system is coupled to the ET system with carefully engineered interactions, the control qubit can then act as a quantum control knob that drives the target system from a trivial initial state to a desired entangled steady state. As an example, we will show that this protocol allows for the efficient generation of highly entangled Dicke states by employing tools available in trapped-ion systems, such as resolved sideband coupling, two-qubit MS interactions, and sympathetic cooling. Moreover, our method does not have explicit constraints on the target system, making it suitable for the dissipative engineering of both spin and boson degrees of freedom. 

The paper is structured as follows: In Sec. \ref{sec_SET}, we review the properties of the single-site ET model in the perturbative regime. In Sec. \ref{sec_EET}, we propose an extended ET model by involving a target quantum system coupled to the control qubit. We demonstrate how this composite system can be used to engineer desired steady-state structures for the target system. We then derive an approximate analytical formalism that quantitatively tracks the dynamics of the system in Sec. \ref{sec_3d}. In Sec. \ref{sec_D}, we propose the major scheme of the paper for engineering steady-state 
$N$-qubit Dicke states through an iterative approach based on the extended ET model. We also show how a similar scheme also allows for the generation of steady-state $N$-boson $W$ states. In Sec. \ref{exr}, we consider an experimental realization of our protocol and evaluate its robustness against noises typically present in state-of-the-art trapped-ion systems. As an example, we simulate the dissipative preparation of a four-qubit Dicke state with two excitations in the presence of noises. Finally, in Sec. \ref{sec_con}, we provide the conclusion and discuss potential outlooks for future research.  

\section{Single-site ET model}\label{sec_SET}

We shall start with a review of the single-site ET model simulated in \cite{so2024trapped}. The Hamiltonian of the model can be decomposed into a Jahn-Teller type of Hamiltonian $H_0$ \cite{o1993jahn} and an electronic coupling term $H_1$ such that $H_{\text{ET}}=H_0+H_1$, with
\begin{equation}
    H_0=\frac{\Delta E}{2}\sigma_z+\frac{g}{2}\sigma_z\left(a+a^\dag\right)+\omega_0a^\dag a,\ \ \ H_1=V\sigma_x,
    \label{ET_H}
\end{equation}
where $\sigma_z,\sigma_x$ are the Pauli operators of a spin-1/2 qubit, and $a,a^\dagger$ are bosonic annihilation and creation operators, respectively.

The damping of the bosonic mode is induced by the linear coupling between the mode and a Markovian environment with an Ohmic spectral density \cite{garg1985effect, Lemmer2018structure}. The dissipative dynamics of the system can be described by a Lindblad master equation \cite{carmichael2013statistical}:
\begin{eqnarray}
 \frac{\partial\rho}{\partial t}&=&-i[H,\rho] + \gamma (\bar{n}+1)\mathcal{L}_{a}[\rho] + \gamma \bar{n} \mathcal{L}_{a^\dagger}[\rho],\nonumber
    \label{eq_master}\\
    \mathcal{L}_{c}[\rho]&=&
    c\rho c^\dagger - \frac{1}{2}\{c^\dagger c,\rho\},
    \label{eq_Lind}    
\end{eqnarray}
where $H=H_{\text{ET}}$, and the average thermal boson number $\bar{n}$ characterizes the temperature of the environment. With $\Delta E =0 $, this model is a variation of the quantum Rabi model with cavity decay \cite{braak2011integrability}. 

The non-interacting Hamiltonian $H_0$ can be rewritten in a quadratic form by introducing two pairs of shifted bosonic operators $b (b^\dag)_\pm \equiv a(a^\dag)\pm \tilde{g}/2$ with $\tilde{g} \equiv g/\omega_0$:
\begin{equation}
H_0=\left|\uparrow\right\rangle\left\langle\uparrow\right|(\Delta E/2+\omega_0b_+^\dag b_+)+\left|\downarrow\right\rangle\left\langle\downarrow\right|(-\Delta E/2+\omega_0b_-^\dag b_-).
\end{equation}
We define the $\sigma_z$ eigenstates of the control qubit as $\ket{D}=\ket{\uparrow}$ (donor), $\ket{A}=\ket{\downarrow}$ (acceptor) and the corresponding two sets of eigenstates of $H_0$ as the donor/acceptor vibronic states, explicitly:
\begin{eqnarray}
    \left|D,n_d\right\rangle&=&\left|D\right\rangle U_{\text{disp}}\left(-\tilde{g}/2\right)\left|n_d\right\rangle,\, E_{n_d}=\Delta E/2+n_d\omega_0,\nonumber\\
    \left|A,n_a\right\rangle&=&\left|A\right\rangle U_{\text{disp}}\left(+\tilde{g}/2\right)\left|n_a\right\rangle,\, E_{n_a}=-\Delta E/2+n_a\omega_0,\nonumber
    \\
    \label{lev_HET}
\end{eqnarray}
where $U_{\text{disp}}(\alpha)=\exp{(\alpha a^\dagger-\alpha^*a)}$ is the displacement operator, and $\left|n_d\right\rangle,\left|n_a\right\rangle$ are Fock states with excitation number $n_d,n_a$, respectively. 

For the purpose of this paper, we focus on the perturbative regime: $V \lesssim\gamma, V\ll \omega_0,V\ll \lambda \equiv g^2/\omega_0$, where $\lambda$ is the reorganization energy of the ET system \cite{schlawin2021continuously,taylor2019theory}. This condition ensures that, with $\gamma$ taken as an effective broadening of the unperturbed energy levels \cite{skourtis1992new}, the effect of $H_1$ on the spectrum of $H_0$ is sufficiently small such that $H_1$ only couples the donor and acceptor eigenstates of the same energy, leading to a resonance condition given by
$\left(n_a-n_d\right)\omega_0=\Delta E$. For simplicity, we consider the case $\Delta E=\omega_0$ and the system initialized in the ground donor state $\left|D,n_d=0\right\rangle$. The only energetically resonant state is therefore $\left|A,n_a=1\right\rangle$.
By further assuming bath temperature being zero ($\bar{n}=0$), the third term in Eq. \eqref{eq_Lind} vanishes. The population in the intermediate state 
$\left|A,n_a=1\right\rangle$ irreversibly decays to the ground acceptor state $\left|A,n_a=0\right\rangle$ through the collapse operator $a$. $\left|A,n_a=0\right\rangle$ is an approximate dark state of the system (for details, see section S5 of the Supplementary Materials in \cite{so2024trapped}) since it is not resonantly coupled to any other eigenstates of $H_{ET}$ and remains invariant under the collapse operator $a$. Therefore, in the steady state, the donor population will be irreversibly transferred to $\left|A,n_a=0\right\rangle$. The above intuition can also be applied to understand the extended ET model to be proposed in the next section. 

\section{Extended ET Model} \label{sec_EET}
As demonstrated above, the single-site ET model enables the dissipative preparation of the acceptor ground state. Entangled steady states can be obtained by coupling additional degrees of freedom from a target quantum system to the control qubit, leading to a composite system that we shall define as the extended ET model. The Hamiltonian of the system takes the following generic form:
\begin{eqnarray}
     H_{\rm tot} &=& \underbrace{H_0+H_{\rm ext}}_{H'_0}+H_{\text{int}},
     \label{ET_ext}
\end{eqnarray}
where $H_0$ is the non-interacting ET Hamiltonian defined in Eq. \eqref{ET_H}. The Hamiltonian involving the target system's degrees of freedom is composed of two parts: \emph{(i)} the interaction Hamiltonian $H_{\text{int}}$ coupling the target system to the control qubit, as well as the donor states to the acceptor states of the ET system ($\bra{A}H_{\text{int}}\ket{D}\neq0$); \emph{(ii)} the external Hamiltonian $H_{\text{ext}}$, which creates additional energy splittings in the eigenstates of $H_0$ without coupling the donor and acceptor vibronic states. The new sets of donor/acceptor vibronic states, which are the unperturbed eigenstates of $H_0'$, can then be written as:
\begin{equation}
\left|D,n_d,i\right\rangle=\left|D,n_d\right\rangle\otimes\left|e_i\right\rangle,\ \ \left|A,{n}_a,i\right\rangle=\left|A,n_a\right\rangle\otimes\left|e_i\right\rangle,
\label{ex_state}
\end{equation}
where $\{\ket{e_i}\}$ represents the set of eigenstates of $H_{\text{ext}}$ labeled by the quantum number $i$. Both $H_{\rm int}$ and $H_{\rm ext}$ may include spin and bosonic degrees of freedom depending on the specific desired entangled state of the target system. 

\begin{figure}[t!]
    \includegraphics[width=1\linewidth]{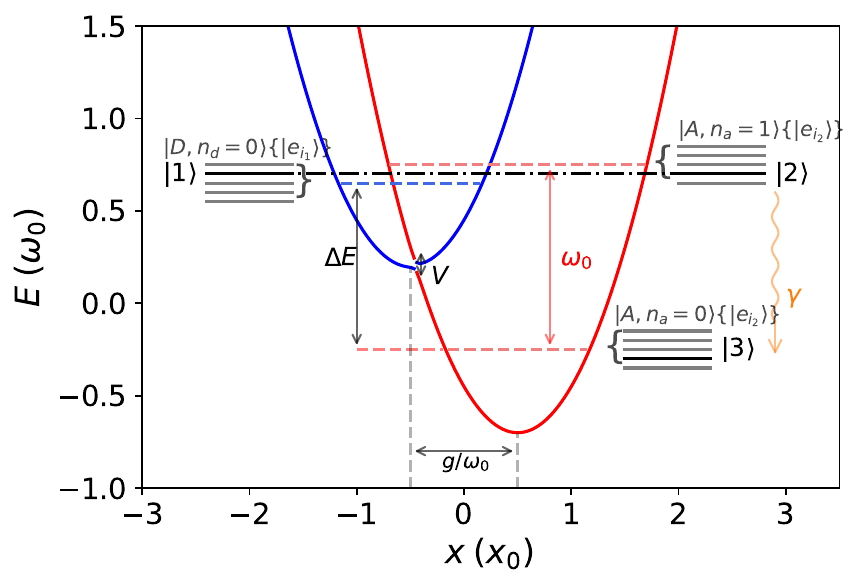}
    \caption{
    \justifying
    Steady-state engineering scheme using extended ET model in Eq. \eqref{ET_ext}. For the purpose of demonstration, we plot the case of $H_{\text{int}}=V\sigma_x$, which is equivalent to $H_{\text{ET}}$ with added pseudo-energy levels. For a specific scheme, one can replace $\sigma_x$ with the actual form of $H_{\text{int}}$ with coupling strength $V$.  The blue and red curves are the donor and acceptor nonadiabatic harmonic potential surfaces plotted with parameters $(V,g,\Delta E)=(0.025,1,0.9)\omega_0$, respectively. The horizontal axis is the position in units of $x_0=\sqrt{2/m\omega_0}$, which is the length scale of a harmonic oscillator with mass $m$ at frequency $\omega_0$. The donor vibronic state $\ket{D,n_d=0}$ and acceptor vibronic states $\ket{A,n_a=1}, \ket{A,n_a=0}$ (ancilla states) are plotted with dashed lines. The three sets of grey lines represent the additional energy levels due to $H_{\text{ext}}$ on each of the vibronic states. Three eigenstates ($\ket{1}, \ket{2}, \ket{3}$) of $H_0'$ that participate in the scheme are plotted in bold black lines. $\ket{1}$ and $\ket{2}$ are energetically resonant (connected by the black dot-dashed lines) and coupled by $V\sigma_x$. $\ket{2}$ then decays to the steady state $\ket{3}$ under the boson dissipation at the rate $\gamma$.}
    \label{fig:scheme}
\end{figure}

The Hamiltonian $H_{\text{tot}}$ enables us to manipulate the steady-state structure of the target system, as illustrated in Fig.~\ref{fig:scheme}. Specifically, suppose the composite system is initialized in a donor vibronic state $\ket{1}\equiv\ket{D,n_d=0,i_1}$. $\Delta E$ is then tuned such that $H_{\text{int}}$ resonantly couples $\ket{1}$ to a single acceptor vibronic state $\ket{2}\equiv\ket{A,n_a=1,i_2}$, assuming no degeneracy. The boson dissipation will sequentially drive the system to an approximate dark state $\ket{3}\equiv\ket{A,n_a=0,i_2}$ that is off-resonant with all donor states, leading to the target steady state $\ket{e_{i_2}}$. The key components of the scheme are the two ancilla vibronic states $\ket{2},\ket{3}$, engineered through the non-interacting ET Hamiltonian $H_0$ in Eq. \eqref{ET_H} on the $\ket{A}$ state of the control qubit. Together with the initial state $\ket{1}$, they form an effective 3-level landscape, which can be found in typical optical pumping schemes, with a decay rate that can be finely tuned through sympathetic cooling parameters. 

 In the degenerate case, where multiple acceptor states are on resonance with $\ket{1}$, it can be shown that the system's population will be transferred to a superposition of these states. This property will be essential to the schemes to be proposed in Sec. \ref{sec_D}.

To ensure the accuracy of energy selection for the intermediate state $\ket{2}$, $H_{\text{int}}$ is required to be maintained as a perturbation of $H_0'$. It is sufficient to set all energy splittings induced by $H_{\text{ext}}$ much smaller than the vibronic energy $\omega_0$, as well as the reorganization energy $\lambda$. At the same time, the interaction strength $V=|\bra{A}H_{\text{int}}\ket{D}|$ is kept much smaller than the splitting of $H_{\text{ext}}$. Let ${\delta E_i}$ be the set of all distinctive energy gaps of $H_{\text{ext}}$. A set of sufficient conditions for the perturbative limit is then given by:
\begin{eqnarray}
    V\lesssim \gamma\ll \min \{\delta E\}, \nonumber \\ \:\max\{\delta E\}\ll \omega_0, \nonumber \\\: \max\{\delta E\}\ll g^2/\omega_0.
    \label{pc}
\end{eqnarray}
Since the condition $g\sim\omega_0$ can be routinely implemented experimentally, this reduces the last two constraints to $\max\{\delta E\}\ll \omega_0$.  

 The steady state of the extended ET system is not necessarily unique and might depend on the initial state. In practical cases, $H_{\text{ext}}$ and $H_{\text{int}}$ typically assume relatively simple forms, leading to the possible emergence of weak or strong symmetries \cite{albert2014symmetries,nigro2019uniqueness,zhang2024criteria}. Therefore, an appropriate initial state $\ket{\psi_i}$ must also be chosen once $H_{\text{int}}$, $H_{\text{ext}}$ are determined.

{Based on the above considerations, engineering a specific target quantum state $\ket{\psi^{\rm ext}_{\text{tar}}}$ of the external/target system using the extended ET model requires a careful design of the set $S_{\text{EET}}=\{H_{\text{int}},H_{\text{ext}},\ket{\psi_i}\}$, where $\ket{\psi_i}=\ket{\psi^{\rm ET}_i}\otimes\ket{\psi^{\rm ext}_i}$ is the initial state of the total system. Given a specific $\ket{\psi^{\rm ext}_{\text{tar}}}$, an appropriate choice of $S_{\text{EET}}$ should satisfy the following conditions simultaneously: 
\begin{enumerate}
    \item It should be experimentally feasible to prepare the initial state of the target system $\ket{\psi^{\text{ext}}_i}$ as one of the eigenstates $\ket{e_{i_1}}$ of $H_{\text{ext}}$, which, therefore, should be a product state to ensure a low experimental overhead in its preparation. Similarly, the ET system $\ket{\psi_i^{\rm ET}}$ has to be prepared close to the donor ground state $\ket{D,n_d=0}$.
    \item The target state has to be either an eigenstate $\ket{e_{i_2}}$ of $H_{\text{ext}}$, or a superposition of degenerate eigenstates $\ket{e_{j,E_k}}$ of $H_{\text{ext}}$ corresponding to a common energy $E_k$. 
    \item  The coupling condition $(\langle \psi_{\text{tar}}^{\rm ext}|\otimes\langle A|)H_{\text{int}}(|D\rangle \otimes|\psi^{\rm ext}_i\rangle)\neq0$ is satisfied so that the state of the target system is flipped whenever the control qubit is flipped from $\ket{D}$ to $\ket{A}$.
    \item The perturbative condition as stated in Eq.~(8) or a relaxed version of it is satisfied, such that the desired eigenstate(s) can be selectively addressed by tuning $\Delta E$ and $\omega_0$, which control the energy matching conditions in the system. 
\end{enumerate}
With an appropriate choice of $S_{\text{EET}}$, the original ET system becomes a control knob for the steady state of the target system. In the following sections, we present concrete example choices of $S_{\text{EET}}$ that lead to entangled steady states of $N$ qubits or bosons, as summarized in Table \ref{tab_sum}. It should be noted that the structure of the target steady state is not arbitrary, because the forms of $H_{\text{int}}$ and $H_{\text{ext}}$ are subject to constraints imposed by the specific experimental platform.
}
\begin{table*}[t!]
\centering
\begin{tabular}{|c|c|c|c|}
\hline
Target state & $H_{\text{int}}$ & $H_{\text{ext}}$  & $\ket{\psi_i^{\rm ext}}$\\
\hline
$N$-qubit $W$ and Dicke states (Sec. \ref{sec_D}) & $J(\sigma^+_0\sum_{i=1}^N \sigma^-_i+ {\rm h.c.})$ & 0  &  $\ket{\downarrow \cdots \downarrow}_N$\\
\hline $N$-boson $W$ states (Sec. \ref{sec_D}), Appendix. \ref{ap_eph})& $J\sum_{i=1}^N(\sigma_0^+b_i+{\rm h.c.})$ & 0 &$\ket{0\cdots 0}_N$ \\
\hline
2$N$-qubit GHZ state (Appendix. \ref{Ap_3}) & $V\sigma_0^x$& $\frac{E_0}{2}(\sigma_0^z+I_0)\sum \nolimits_{i=1}^{2N}\sigma^z_i+
    \frac{k}{2}\left(\prod \nolimits_{i=1}^N\sigma^x_{i} +\prod \nolimits_{i=N+1}^{2N}\sigma^x_{i} \right)$ &  $\ket{\uparrow \cdots \uparrow}_{2N}$\\
\hline
\end{tabular}
\caption{Choices of $S_{\text{EET}}$ for the schemes proposed in this work. Here instead of $\ket{\psi_i}$ of the composite system, we list the initial state of the target system $\ket{\psi_i^{\rm ext}}$,  since in the ideal case the corresponding $\ket{\psi_i}$ can be obtained by making a direct product with the ground donor state $\ket{\psi^{\rm ET}_i}=\ket{D,n_d=0}$. The index $0$ in the Hamiltonians denotes the control qubit, while the indices $1\sim N(2N)$ denotes the target qubits or bosons.}
\label{tab_sum}
\end{table*}


The fact that the damped bosonic mode acts as a non-Markovian bath for the control qubit plays a crucial role in our protocol. This is evident when comparing the single-site ET model with the system studied in Ref. \cite{huelga2012non}, which consists of two spins linearly coupled to two independently damped bosonic modes with coupling strengths $g_1, g_2$ and a common damping rate $\kappa$. In the latter, it has been shown that non-Markovianity is essential for generating steady-state entanglement in their system. By identifying $\ket{\uparrow}, \ket{\downarrow}$ with $\ket{\uparrow\downarrow}, \ket{\downarrow\uparrow}$ and setting $\Delta E = 0$, $g = g_1 - g_2$, and $\gamma = \kappa$, the single-site ET model described by Eqs.\eqref{ET_H} and \eqref{eq_Lind} maps directly onto their system in the strongly non-perturbative regime ($V \sim g \sim \omega_0$). Although the additional constraint $V \sim g \sim \omega_0$, necessary for generating a steady-state Bell singlet in Ref. \cite{huelga2012non}, does not directly apply to our case, the non-Markovian condition $g \gg \gamma$ is naturally satisfied by the perturbative condition in Eq.~\eqref{pc}. In the Markovian regime, where $\gamma \gg g$, the dissipative dynamics of both the ET and extended ET models reduce to pure dephasing on the control qubit with a collapse operator $\sigma_z$, rendering the transfer dynamics described above invalid and thereby preventing our protocol from functioning.

\subsection{Reduced 3-level Formalism for Extended ET Model}\label{sec_3d}

Under the conditions in Eq. \eqref{pc}, the dynamics of the ET system can be approximated by a two-level model~\cite{schlawin2021continuously}. Previously, the quantum jump term $a \rho a^\dagger$ and the fact that the bosonic components of the eigenstates can be highly displaced states instead of Fock states have been ignored. Here, we take these effects into account by considering a reduced 3-level system in the subspace spanned by $\{\ket{1},\ket{2},\ket{3}\}$ defined above. This allows us to track the dynamics described by the master equation in Eq. \eqref{eq_Lind} with $H=H_{\rm tot}$ and $\bar{n}=0$ more accurately. Denoting $E_0$ as the energy difference between $\ket{e_{i_1}}$ and $ \ket{e_{i_2}}$, the resonance between $\ket{1}$ and $\ket{2}$ requires $\Delta E =\omega_0-E_0 $.  Eq. \eqref{eq_master} 
gives a set of 8 independent first-order linear differential equations for all density matrix elements $\rho_{ij}$  with $i,j=1,2,3$. Since the transition from $\ket{1}$ to $\ket{3}$ is energetically suppressed ($\omega_0\gg V$), and $\ket{2}$, $\ket{3}$ are only connected by the collapse operator, the coherences between these two pairs of states are negligible. This enables a further approximation that $\text{Re}\left[\rho_{23}\right],\;\text{Im}\left[\rho_{13}\right]\rightarrow0 $. Under this approximation, a closed set of three equations consisting of $\text{Im}\left[\rho_{12}\right],\ \rho_{11},\ \rho_{22}, $ can be obtained, which is sufficient for our purpose of studying the population dynamics of $\ket{1},\ket{2},$ and $\ket{3}$:  
 \begin{equation}
     \partial_t\vec{\rho}=M\vec{\rho},\ \ \vec{\rho}=\left[\begin{matrix}\rho_{11}\\\text{Im}[\rho_{12}]\\\rho_{22}\\\end{matrix}\right],\ \ M=\left[\begin{matrix}0&-2V_{\text{e}}&0\\V_{\text{e}}&-g^\prime\gamma&-V_{\text{e}}\\0&2V_{\text{e}}&-\gamma\\\end{matrix}\right],\label{dif_3} 
 \end{equation}
 where $V_{e}
=-V \tilde{g} e^{(-\tilde{g}^2/2)} \bra{e_{i_2}}H_{\text{int}}\ket{e_{i_1}}$ is the effective Rabi frequency that depends on the Franck-Condon factors, and $g^\prime=\frac{1}{2}\left(1+\tilde{g}^2\right)$. With $\gamma,V_{\text{e}}\neq0$, $\text{det}M=-2V_e^2\gamma\neq0,$ making the null space of $M$ contain only the null vector ($\vec{\rho}_{ss}=\vec{0}$). Therefore, in the steady state, there is no population remaining in $\ket{1}$ and $\ket{2}$ of the reduced system, which implies that their populations are completely transferred to state $\ket{3}$. We emphasize that in the full model, the actual steady-state population of $\ket{3}$ will always be less than $1$ due to the small coupling between $\ket{1}$ and $\ket{3}$. 

\begin{figure}[t!]
    \includegraphics[width=1\linewidth]{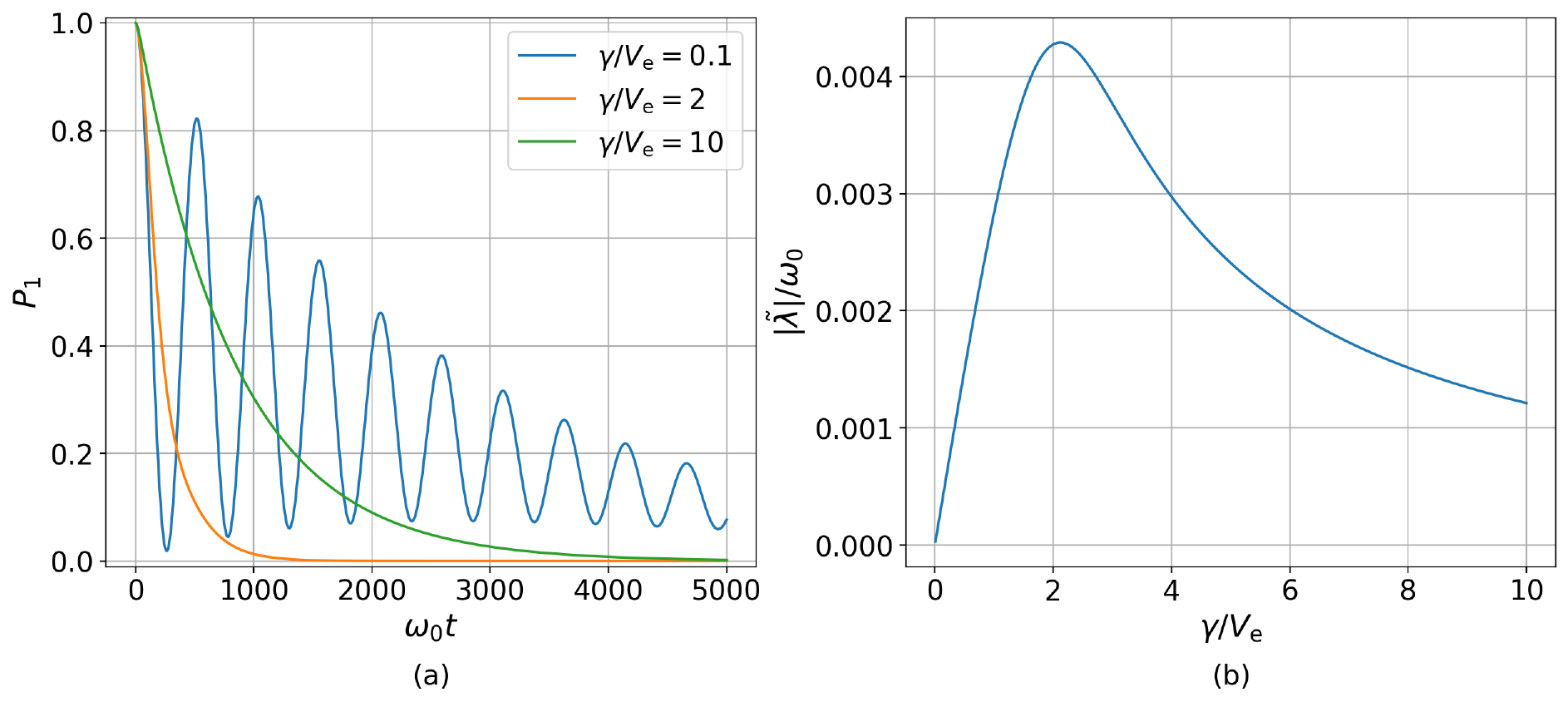}
    \caption{\justifying
    (a) Time evolution of the population in state $\ket{1}$ with $\gamma/V_{\text{e}}=0.1,2,10$. (b) Plot of the normalized maximum negative eigenvalue $|\tilde{\lambda}|/\omega_0$, which estimates the transfer rate as a function of $\gamma/V_{\text{e}}$. The parameters used are $(V,g) = (0.01,1)\omega_0$. In the weak dissipation regime with $\gamma\ll V_{\text{e}}$, $|\tilde\lambda|\propto\gamma$, while $|\tilde\lambda|\propto1/\gamma$ in the strong dissipation regime with $\gamma\gg V_{\text{e}}$. The optimal transfer is located at $\gamma\sim2V_\text{e}$.}
    \label{fig:lambda}
\end{figure}

The analytical solution of Eq. \eqref{dif_3} can be obtained by solving the eigenvalues and eigenvectors $M$, but the exact form is cumbersome due to the complexity of the cubic characteristic equation. We plot a set of typical solutions in Appendix \ref{Ap_A} and compare them with the dynamics obtained by numerically solving the full master equation to check the validity of the approximations. 

The eigenvalues of $M$ give information on the typical timescale for generating the desired steady state. Let us denote the three eigenvalues of $M$ as $\{\lambda_i\}$ with $ \text{Re}[\lambda_i<0]$. The magnitude of 
$\tilde{\lambda}=\max\{\text{Re}[\lambda_i]\}$ determines the rate of the slowest dynamical process and, therefore, gives an estimation of the transfer rate to the steady state. The eigenvalue $\tilde{\lambda}$ is determined by the coherent timescale $1/V_{\text{e}}$ and dissipative timescale $1/\gamma$ of the system. In Fig. \ref{fig:lambda}, we plot typical dynamics of the system\footnote{Simulation data in this study are available in a public repository: \cite{zhu_2025_15794550}.} to demonstrate how $|\tilde{\lambda}|$ changes with the ratio $\gamma/V_{\text{e}}$. An optimal transfer is located at $ \gamma \sim 2V_{\text{e}}$, verifying that our model is consistent with Refs. \cite{schlawin2021continuously,so2024trapped}.  

It is also necessary to take into account a more experimentally realistic scenario with finite boson temperature, where the initial state of the bosonic mode is a displaced thermal state with an average excitation $n_0$, and the bath has a non-zero temperature ($\bar{n}\neq0$). For convenience, we assume the system is initially in thermal equilibrium with the bath such that $n_0=\bar{n}$. This assumption will also be used in later sections of the paper to simplify the analysis. The reduced model fails to capture the exact dynamics under such conditions, but the overall behavior of the system can be understood in a similar manner as the case of $\bar{n}=0$\footnote{It has been shown in the Supplementary Materials Section S5 of Ref.\cite{so2024trapped} that the acceptor steady state of the ET model is a displaced thermal state characterized by $\bar{n}$.}. We note that the steady-state population of $\ket{3}$ decreases as $\bar{n}$ increases, resulting in a reduction in the fidelity of the target state. This effect can be mitigated by increasing the energy splitting $ \Delta E$ to an optimal level, where higher-order resonances are induced, while maintaining a sufficiently strong, albeit smaller, vibronic coupling. We provide a more detailed discussion of the mechanism in Appendix \ref{ap_FT}. This technique will be employed to enhance the fidelity of the schemes proposed in the later sections.


\section{Steady-state $N$-qubit Dicke state}\label{sec_D}

In this section, we develop the major scheme of this paper:  dissipative generation of a $N$-qubit Dicke state with $m$ controllable excitations. The target state is denoted as $\ket{W_N^m}$, which is a simultaneous eigenstate of total angular momentum operators $J^2$ and $J_z$. When $m=1$, the target state is usually referred to as the $W$ state. The $W$ states are particularly important since the  entanglement contained is robust against particle loss \cite{dur2000three}. For example, there are proposals to realize steady-state three-qubit $W$ states of different chiralities and {$N$-qubit $W$ states as the ground state of a quantum spin chain in trapped-ion systems ~\cite{cole2021dissipative,cormick2013dissipative}} . 
The more generic family of Dicke states also plays a key role in many areas of quantum computing, including topological data analysis \cite{berry2024analyzing} and quantum networking \cite{prevedel2009experimental,miguel2020delocalized}. In the following, we outline protocols to create $N$-qubit $W$ and Dicke states.

\textbf{$W$ Spin States} - We first demonstrate a scheme for generating $N$-qubit $W$ states. Let us denote the control qubit with index $0$ and add $N$ target qubits to the ET model. We utilize spin-spin interactions between the control qubit and each of the target qubits such that $H_{\text{ext}} = 0$, and
$ H_{\text{int}} = \sigma^+_0\sum_{i=1}^N J_i\sigma^-_i+ {\rm h.c.}$
The total Hamiltonian then takes the following form:
\begin{equation}
    H_{\rm tot} = \frac{\Delta E}{2}\sigma^z_0+ (\sigma^+_0\sum_{i=1}^NJ_i\sigma^-_i+ {\rm h.c.})+ \frac{g}{2}\sigma_0^z\left(a+a^\dag\right)+\omega_0a^\dag a.
    \label{Hd1}
\end{equation}

Let us consider the system initialized in a donor vibronic state $\ket{\psi_i}=\left|D,n_d=0\right\rangle\otimes\ket{\downarrow\hdots \downarrow}$, such that all the target spins are in $\ket{\downarrow}$ state. Since $[H_{\text{int}}, J_z]=0$, the total spin excitation is preserved in the system. Our analysis of the extended ET model in the previous section shows that the control qubit will be pumped to $\ket{A}$ state if the perturbative limit holds. Specifically, the perturbation conditions in Eq. \eqref{pc} reduces to:
\begin{equation}
J_{i}\sim\gamma\ll\omega_0,
\label{pc2}
\end{equation}
assuming $\omega_0\sim g$. At the same time, the donor state excitation will be deterministically delivered to the target qubits by $H_\text{int}$. Hence, we expect a superposition of single-excitation states in the steady state.  

To quantitatively track the dynamics, under the condition \eqref{pc2}, it is sufficient to consider only the other $N$ excitation-conserving states coupled to $\ket{\psi_i}$ by $H_{\text{int}}$, which are all on resonance if we set $\Delta E = n\omega_0, n\in \mathbb{N}$. Specifically, if $n=1$, these states are given by $\{\ket{k^+}\} = \{ \ket{A,n_d=1} \otimes \ket{\{k\}}\}$ with $k=1,\hdots, N$, where $\ket{\{k\}}=\ket{\downarrow_1\dots\downarrow_{k-1}\uparrow_k\downarrow_{k+1}\dots\downarrow_N}$ denotes a $N$-qubit product state with the $k$-th qubit in $\ket{\!\uparrow}$ and all other qubits in $\ket{\!\downarrow}$. In this basis, the total Hamiltonian $H_{\rm tot}$ takes a simple form:
    \[ H_{\rm tot} = 
\begin{bmatrix}
0 & V'_1 & V'_2 & \cdots & V'_N \\
V'_1 & 0 & 0 & \cdots & 0 \\
V'_2 & 0 & 0 & \cdots & 0 \\
\vdots & \vdots & \vdots & \ddots & \vdots \\
V'_N & 0 & 0 & \cdots & 0
\end{bmatrix},
\]
where the effective coupling strength $V_k'$ is given by $-J_k \tilde{g} e^{(-\tilde{g}^2/2)} $.
$H_{\rm tot}$ has only two non-zero eigenvalues, $\pm V_s=\pm\sqrt{\sum\nolimits_{k=1}^N {V'_k}^2}$ with the 
corresponding eigenvectors: 
\begin{equation}
    \ket{\pm V_s} = \frac{1}{\sqrt{2}}\left(\pm\ket{1}+\frac{1}{V_s}\sum\nolimits_{k=1}^N {V'_k}\ket{k^+}\right).
\end{equation}
The other eigenvectors have zero eigenvalues, and they have no overlap with $\ket{1} = \ket{\psi_i}$. Hence, the eigenbasis decomposition of $\ket{1} $ only involves $\ket{\pm V_s}$. Following the formalism introduced in Section \ref{sec_3d} by identifying 
\begin{eqnarray}
    \ket{2}&=&  \ket{A,n_d=1}\otimes \frac{1}{V_s}\sum\nolimits_{k=1}^N {V'_k}\ket{\{k\}},\;\;    \nonumber\\
    \ket{3}&=& \ket{A,n_d=0}\otimes \frac{1}{V_s}\sum\nolimits_{k=1}^N {V'_k}\ket{\{k\}},
    \label{pe_23}
\end{eqnarray}
the dynamics of the system can then be described by Eq.~\eqref{dif_3} with the effective Rabi frequency $V_{\text{e}}=V_s$. In the steady state, the system approaches $\ket{3}$ with entanglement within the $N$ target qubits. Specifically, if the couplings to the control qubit $J_i$ are set identically as $J$, the steady state becomes the target state $\ket{W}=\frac{1}{\sqrt{N}}\sum\nolimits_{k=1}^N \ket{\{k\}}$.

{It shall be noted that, a Hamiltonian with a form very similar to Eq.~\eqref{Hd1} has been used in Ref.~\cite{cormick2013dissipative} for the dissipative generation of ground states of a spin-chain, including the $N$-qubit $W$ state.  However, the underlying physical mechanism of this protocol differs significantly from the one we have just described. In their setup, the $N$ target spins are coupled to a damped bosonic mode with fine-tuned, spin-dependent coupling strengths, forming a damped spin-boson chain (DSBC). When combined with an isotropic XY Hamiltonian with local transverse fields on the edge, the spin-boson couplings and the bosonic damping effectively realize a collective jump operator composed of single-spin operators.  The DSBC system is then dissipatively pumped to an $N$-qubit $W$-like state if an appropriate boson frequency is chosen. We denote this state preparation method as the DSBC protocol in the following. Compared to the DSBC protocol, our proposed scheme offers several advantages: (1) The optimal operating regime for the DSBC protocol requires $g \sim \kappa \ll J$, where $g$ is the spin-boson coupling strength, $\kappa$ is the bosonic dissipation rate, and $J$ denotes the strength of the XY spin-spin interaction. Consequently, the gate time of the DSBC protocol, given by $\tau \sim \kappa / |g|^2$, is fundamentally limited by the maximum achievable value of $J$. In contrast, in our scheme, the gate time is primarily constrained by the spin-boson coupling $g$ (see detailed discussion in the next section). In typical trapped-ion setups, $g\gg J$, which allows for shorter gate times. (2) The DSBC protocol is inherently sensitive to boson heating, whereas in our approach, this effect can be effectively mitigated by optimizing the control qubit energy splitting $\Delta E$ to protect the fidelity of target state (see Appendix \ref{ap_FT} for further details). (3)  The DSBC protocol does not easily generalize from $W$ states to Dicke states with more than one spin excitation, whereas our method exhibits greater flexibility in this regard, as we will show in the next few paragraphs.} 

\textbf{Dicke Spin States} - The above protocol can naturally be extended to realize a Dicke state with $M$ excitations in the steady state by involving a chain of $N+M$ ions with $M$ control qubits, each coupled to an independent damped bosonic mode, and $N$ target qubits. In this case, the total Hamiltonian takes the form:
\begin{eqnarray}
     H_{\rm tot} &=& \sum_{i=1}^{M}\left[\frac{\Delta E}{2}\sigma^z_{0,i}+\frac{g}{2}\sigma_{0,i}^z(a_i+a_i^\dag)+\omega_0{a_i^\dag a_i}\right] \nonumber
     \\&\quad&+  \sum_{i=1}^M\left[\sigma^+_{0,i}\sum_{j=1}^N J_{i,j}\sigma^-_j+ {\rm h.c.}\right].
    \label{Hd2}
\end{eqnarray}
In this generalization, the $M$ control qubits are all initialized in $\left|D,n_d=0\right\rangle$, so that, in the steady state, the $M$ spin excitations will be transferred to the $N$ target qubits, which are initially prepared in the $\ket{\!\downarrow}$ state, forming a Dicke state with $M$ excitations. 

\begin{figure}[t]
\centering
\includegraphics[width=\columnwidth]{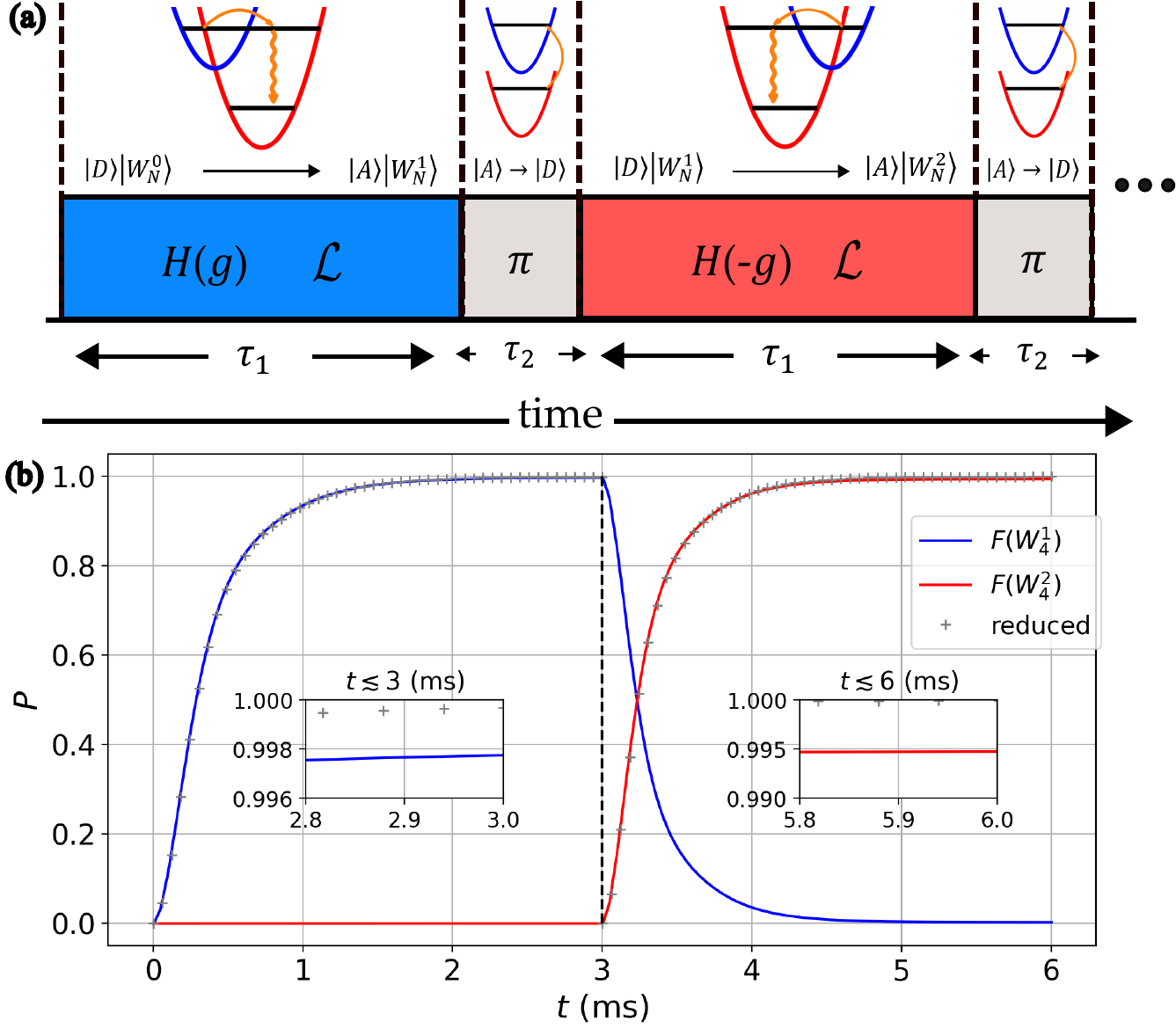}
\caption{\justifying (a). Demonstration of the hybrid scheme for generating $\ket{W_N^m}$. In each pumping step with gate time $\tau_1$, the Hamiltonian in Eq. \eqref{Hd1} is applied with sympathetic cooling to pump one spin excitation from the control qubit to the target qubits. The sign of $g$ is alternated between two consecutive iterations. In each repumping step, a $\pi$ pulse with gate time $\tau_2$ is applied to repump the control qubit back to the $\ket{D}$ state. (b). Population dynamics of the hybrid scheme for generating $\ket{W_4^2}$. The blue and red solid lines are the fidelities of $\ket{W_4^1}$ and $\ket{W_4^2}$ obtained by numerically integrating Eq. \eqref{eq_master}, with $\tau_1=3$ ms and $\tau_2=1\;\mu$s. The black dashed line marks the time at which the repumping $\pi$ pulse is applied. The grey cross points are the analytical results obtained by combining Eqs. \eqref{dif_3} and \eqref{Ve_m}.}
\label{fig:schemeA}
\end{figure}

\begin{figure}[t]
\centering
\includegraphics[width=\columnwidth]{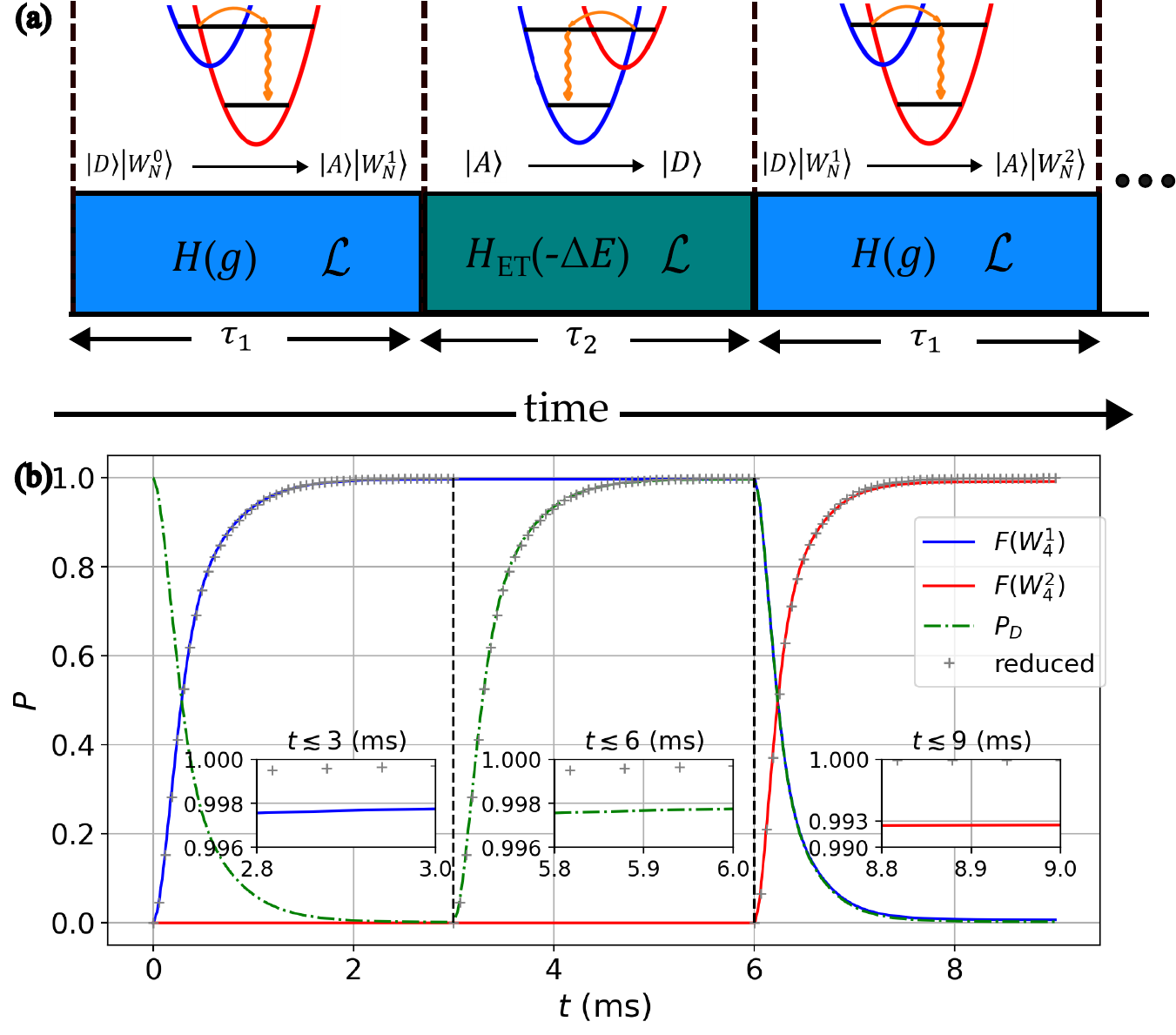}
\caption{\justifying (a). Demonstration of the fully dissipative scheme for generating $\ket{W_N^m}$. In each pumping step with gate time $\tau_1$, the Hamiltonian in Eq. \eqref{Hd1} is applied with sympathetic cooling to pump one spin excitation from the control qubit to the target qubits. In each repumping step, the Hamiltonian in Eq. \eqref{ET_H} with the sign of $
\Delta E$ reversed, and sympathetic cooling is applied to repump the control qubit back to $\ket{D}$ state with gate time $\tau_2$. (b). Population dynamics of the fully dissipative scheme for generating $\ket{W_4^2}$. The blue and red solid lines are the fidelities of $\ket{W_4^1}$ and $\ket{W_4^2}$ obtained by numerically integrating Eq. \eqref{eq_master}, with $\tau_1=\tau_2=3$ ms. The time interval bounded between the two black dashed lines ($3\sim6$ ms) is the repumping step. To track the repumping, we also plot the population of $\ket{D}$ state with a green dashed line. The grey cross points are the analytical results obtained by combining Eqs. \eqref{dif_3} and \eqref{Ve_m}.}
\label{fig:schemeB}
\end{figure}

As shown in the Hamiltonian in Eq. \eqref{Hd2}, this protocol requires $M$ independently cooled bosonic modes being coupled to the control qubits, and $M\times N$ pairs of carefully engineered coupling between control and target qubits of the same strength, leading to a considerable experimental overhead. Here we propose an experimental protocol based on $M$ sequential applications of the Hamiltonian in Eq. \eqref{Hd1}, which is more experimentally feasible, as it relies on a single control qubit (see details of the experimental realization in the next section). In this scheme, each iteration $m$ pumps one spin excitation from the single control qubit to the $N$ target qubits, dissipatively preparing $\ket{W_N^{m+1}}$ state starting from the $\ket{W_N^m}$ (see Figs.~\ref{fig:schemeA}(a) and \ref{fig:schemeB}(a)). After the excitation is transferred,
the control qubit is re-initialized to $\ket{D}$ before the next iteration. Since only the pumping step is applied in the last iteration, $M$ pumping steps and $M-1$ re-initialization steps are required to generate the target state. 
By assuming identical couplings between the control qubit and each target qubit ($J_i=J$), following a similar derivation as for $V_s$ in the case of $W$ states, we identify the effective Rabi frequency for the $m$-th pumping step ($\ket{W_N^m}\rightarrow\ket{W_N^{m+1}}$) as:
\begin{equation}
    V_{\text{e}}^m = \sqrt{(N-m)(m+1)}J\tilde{g}e^{-\tilde{g}^2/2}.
    \label{Ve_m}
\end{equation}
Using this expression in the reduced dynamics described in Eq. \eqref{dif_3}, we can analytically track the dynamics of each pumping process. 

The application of this scheme can be either hybrid or fully dissipative, depending on the choice of the re-initialization mechanism. In the former case (Fig. \ref{fig:schemeA}(a)), after each pumping step, a coherent $\pi$ pulse is applied to the control qubit. In the subsequent pumping step, the sign of $g$ is reversed such that the system remains in the approximate ground donor vibronic state when the next pumping starts. Alternatively, as shown in Fig. \ref{fig:schemeB} (a), after each pumping step, the dissipative evolution described in Eq. \eqref{ET_H} with $H=H_{ET}$ and $-\Delta E$ can be used to dissipatively re-initialize the control qubit from the ground acceptor vibronic state back to the ground donor state. 
To test the validity of the proposal, we simulate the dissipative generation of the $\ket{W_4^2}$ state by numerically integrating Eq. \eqref{eq_master} with zero boson temperature ($\bar{n}=0$) for both hybrid and fully dissipative schemes \cite{JOHANSSON20121760}. The Hamiltonian for each pumping process is given by Eq. \eqref{Hd1} with $(\Delta E,g,J)=(1,1,0.025)\omega_0$ with $\omega_0=2\pi\times20\;\text{kHz}$. The dissipation rate is set to $\gamma=2V_{\text{e}}^m\gtrsim0.05\omega_0$ for each pumping process to optimize the transfer rate, with $V _e$ given by Eq. \eqref{Ve_m}. For the hybrid scheme, a $\pi$ pulse of the form $\Omega_\pi\sigma_0^x$ is applied to the control qubit for repumping with the $\pi$ time of $1\;\mu s$. For the fully dissipative scheme, we apply $H_{\text{ET}}$ given by Eq. \eqref{ET_H} with $(\Delta E,g,J)=(-1,1,0.05)\omega_0$ for repumping. The corresponding population dynamics of the intermediate state $\ket{W_4^1}$ and target state $\ket{W_4^2}$ are plotted in Figs.~\ref{fig:schemeA}(b) and \ref{fig:schemeB}(b) for the two schemes. $\ket{W_4^1}$ is obtained in $3$ ms with $99.8\%$ fidelity, while the $\ket{W_4^2}$ is obtained in $6$ ms with $99.5\%$ fidelity for the hybrid scheme and $9$ ms with $99.3\%$ fidelity for the fully dissipative scheme. The fundamental limitation of the fidelity are discussed in Appendix \ref{ap_FT}.

The hybrid scheme is more favorable in terms of practical purposes, as the fully dissipative scheme requires a longer re-initialization time $(m-1)\tau$, with $\tau$ being the re-initialization time of a single iteration. 
For this reason, we will focus on a realistic experimental implementation using the hybrid scheme in the next section. 

\textbf{$W$ Boson States} - The protocol proposed here can also be employed for the dissipative generation of a $N$-boson $W$ state ($m=1$) if, instead of $N$ qubits, a target system consisting of $N$ bosonic modes is coupled to the ET sub-system. We shall substitute the spin-spin interaction part in Eq. \eqref{Hd1} with a set of Jaynes-Cummming spin-boson couplings, experimentally realized by applying a laser tone resonant with the motional modes \cite{leibfried2003singleion}. In this case, the Hamiltonian becomes:
\begin{equation}
      H_{\rm tot} = \frac{\Delta E}{2}\sigma^z_0 + \frac{g}{2}\sigma_0^z\left(a+a^\dag\right)+\omega_0a^\dag a +J\sum_{i=1}^N(\sigma_0^+b_i+{\rm h.c.}),\label{H_eph}
\end{equation}
where $b_i^\dagger,b_i$ are bosonic operators of the $i$-th target mode. 
With the system prepared in $\ket{\psi_i}=\ket{D,n=0}\ket{0\hdots0}$ (all $N$ target modes are in the ground state), a single pumping step can transfer the spin excitation to the $N$ bosons to obtain the target entangled bosonic state. However, it is not possible to straightforwardly extend this scheme to prepare bosonic $W_N^m$ Dicke states since, differently from $1/2$ spins, bosons have infinitely many excited levels leading to a phonon steady state that is not in a well-defined Dicke excitation manifold.
In this case, we note that the fidelity of the $W$ boson state is largely affected by how close to zero temperature the initial state $\ket{\psi_i}$ can be prepared, especially when $N>2$. In Appendix \ref{ap_eph}, we show the fidelity of generating a two-boson triplet state ($W_2^1$).

The general protocol we proposed can be applied in distinctive ways. In Appendix \ref{Ap_3}, we also include a deterministic scheme for preparing a $2N$-qubit GHZ state utilizing two $N$-spin coupling and $2N$ spin-spin interactions. This scheme is theoretically plausible but may involve more experimental steps for its realization.

{\section{Proposed Experimental Setup}}\label{exr}

The existing tools used for quantum simulation with trapped ions are readily sufficient to realize the proposed schemes. The terms in the total Hamiltonian in Eq. \eqref{Hd1} that govern the coherent dynamics include the carrier and first-order motional sideband transition drives, the MS spin-spin interaction and spin-boson coupling generated by the spin-dependent force of two symmetrically detuned first-order motional sideband transition drives, and the effective spin-hopping interaction from combining the MS interaction with a large transverse field induced by either an additional common offset to the detunings from the first-order motional sidebands or a simultaneous carrier transition drive \cite{monroe2021programmable, schneider2012simulation}. All of these interactions are well-studied and allow for precise experimental control. By inserting spectator coolant ions in the chain, we can implement continuous resolved sideband cooling on them to generate the dissipator used in our proposal. Therefore, in this section, we shall focus on possible ways to experimentally realize the interactions that are not readily available within the traditional simulation tools, followed by a discussion about the potentially associated noises in the implementation. We end this section by presenting a numerical simulation of the dissipative generation of a Dicke state $\ket{W_4^2}$ as an example for the application of the scheme in realistic experimental conditions. 

\subsection{Selective Pairwise Spin-Hopping Interaction}

A possible scheme for inducing the selective pairwise spin-hopping interaction between the control qubit and $N$ target qubits without any spin-spin interactions among the target qubits themselves, $H_{\rm int}=\sigma^+_0\sum_{i=1}^N J_i\sigma^-_i+ \rm{h.c.}$ is to simultaneously apply two MS laser drives and a large transverse field generated by individual-addressing laser beams. The general idea is that the first MS interaction will generate the coupling between the control qubit and target qubits plus unwanted interactions among target qubits, while the second MS drive has the role of canceling out these interactions.
\par Consider the effective MS spin-spin interaction from precisely controlled bichromatic individual-addressing light fields propagating along a trap axis. With qubit-specific on-resonant Rabi coupling strength $\Omega_i$, spin and motional phases $\phi_{i}$ and $\psi_{i}$, and frequencies of $\pm \mu$ in the dispersive regime ($\mu-\omega_m\gg\eta_{im}\Omega_i \medspace,\, \forall m$, where $\eta_{im}$ is the Lamb-Dicke parameter for ion $i$ in the motional mode $m$ with frequency $\omega_m$ on a chain of $N+1$ ions), the Hamiltonian for the interaction is given by
\begin{align}
    H_\text{MS} = \sum_{\substack{i, j=0\\i<j}}^N J_{ij}\cos{(\psi_i - \psi_j)} \sigma_{\phi_i}\sigma_{\phi_j},
    \label{H_MS}
\end{align}
where the spin operator $\sigma_{\phi_i}\equiv\cos{(\phi_i)}\sigma_x^i+\sin{(\phi_i)}\sigma_y^i$, and the spin-spin coupling strengths are given by $J_{ij}=\Omega_i\Omega_j\sum\limits_{m}\frac{\eta_{im}\eta_{jm}\omega_{m}}{\mu^2-\omega_{m}^2}$. With the addition of a strong transverse field $B\sigma_i^z$ in the limit of $J_{ij}\ll B \ll \mu - \omega_m \medspace,\, \forall m$, which only causes a global shift in the energy landscape of the uncoupled ET Hamiltonian $H_0$ in Eq. \eqref{ET_H}, the spin-spin interaction in Eq. \eqref{H_MS} can be approximated as an effective spin-hopping Hamiltonian that conserves the number of spin \cite{monroe2021programmable}, 
\begin{align}
    H_J = \sum_{\substack{i, j=0\\i<j}}^N J_{ij}\cos{(\psi_i - \psi_j)}(\sigma_i^+\sigma_j^- + \text{h.c.}).
    \label{H_pm}
\end{align}
This Hamiltonian can be explicitly decomposed into
\begin{align}
    H_J = & \medspace \sigma^+_0\sum_{i=1}^N J_i\cos{(\psi_0 - \psi_i)}\sigma^-_i+ \rm{h.c.} \\ \nonumber & \medspace + \sum_{\substack{i, j=1\\i<j}}^N J_{ij}\cos{(\psi_i - \psi_j)}(\sigma_i^+\sigma_j^- + \text{h.c.}).
\end{align}
Similarly, we shall introduce another MS interaction from the light fields propagating along a trap axis orthogonal to the previous MS drive with primes on the laser parameter notations (see Fig. \ref{fig, W42_ex}(a)). However, these light fields are absent on ion $i=0$ and applied to all the other ions. The total Hamiltonian resulting from the two MS drives with a strong transverse field becomes $H_\text{int}=H_J+H_{J'}$, where the Hamiltonian of the second MS drive is
\begin{align}
    H_{J'} = \sum_{\substack{i, j=1\\i<j}}^N J_{ij}'\cos{(\psi_i' - \psi_j')}(\sigma_i^+\sigma_j^- + \text{h.c.}).
\end{align}
We thus can eliminate the undesired spin-spin interactions in the terms that do not involve qubit $i = 0$ by adjusting the light field parameters such that the conditions of $\psi_i-\psi_j = \psi_i' - \psi_j' \pm k\pi$ for $k$ is an odd integer, $J_{ij}=J_{ij}'$, and $\phi_i = \phi_i'$ are met. The last condition is not necessarily strict, but is intended for experimental convenience. The resulting Hamiltonian with $\psi_0 - \psi_i = 2\pi k$ for any integer $k$ is then the selective pairwise spin-hopping Hamiltonian, $H_\text{int}=\sigma^+_0\sum_{i=1}^N J_i\sigma^-_i+ \rm{h.c.}$, used in our proposed scheme for generating entangled spin states.

\subsection{Experimental Noise Sources}
Three types of noises must be taken into account due to possible experimental imperfections in realizing the above pairwise interactions in our proposed setup. 

\textbf{Non-excitation-conserving couplings.} These couplings refer to the off-resonance terms, $J_{ij}\sigma_i^+\sigma_j^++\text{h.c.}$, which should be sufficiently suppressed by the global transverse field of strength $B$. The residual contributions directly deteriorate the final fidelity as they tend to connect components of $\ket{W_N^k}$ to states orthogonal to $\ket{W_N^{k+1}}$ in each pumping step. Further increasing $B$ to ensure the $B\gg J$ condition suppresses this error. 

\textbf{Imbalance in pairwise coupling strengths.} To obtain a true Dicke state, we require all $J_{0,j}=J$, which cannot be perfectly controlled in a realistic setup. Therefore, to quantify its effect on the target state fidelity, we define the imbalance as $\delta J=\underset{\{i,j\}}{\max}{|J_{0,i}-J_{0,j}|}$. A non-zero $\delta J$ introduces unequal weights to the components of the target state during each pumping step, leading to deviations from the ideal Dicke state. Furthermore, with the presence of non-excitation-conserving couplings of the form $\sim\sigma^\pm_i\sigma^\pm_j$, dissipation can no longer freeze the system in a steady state. Instead, the state corresponding to the target qubits will oscillate around the desired steady state. Nonetheless, when the imbalance is sufficiently small ($\delta J\lesssim0.01J)$, it will not have a significant effect on the final fidelity. 

\textbf{Residual couplings between target qubits.} With the presence of non-excitation-conserving or imbalanced couplings, the residual couplings, $\sigma_i^+\sigma_j^-+\text{h.c.}$ for $i,j\neq0$, can also induce oscillations since the desired steady state is not invariant under these terms. The effect of the residual terms is negligible if $J_{\text{res}}\lesssim0.01J$, where $J_{\text{res}}\equiv\underset{i,j\neq0}{\max}J_{ij}$.

In addition to the three factors above, we must enforce the perturbative limit for the extended ET model in each pumping step such that nearly all the population of the control qubit is transferred from $\ket{D}$ to $\ket{A}$. Along with the condition in Eq. \eqref{pc2}, we also require 
\begin{equation}
    J\sim\gamma\ll\underset{n\leq n_c,\;n\in\mathbb{N}}{\min}{|4B-n\omega_0|},
    \label{pc3}
\end{equation}
where $4B$ is the energy difference induced by the non-excitation conserving terms $\sigma_i^\pm\sigma_j^\pm$ under the transverse field $BJ_z$, $n_c$ is an integer cutoff such that the vibronic state overlap, $\tilde{g}^{n'}e^{-\tilde{g}^2/2} /\sqrt{n'!}$ for $n'=n_c+\Delta E/\omega_0$, is sufficiently small. This extra condition ensures that the energy barrier induced by the global transverse field to suppress the non-excitation-conserving terms cannot be compensated by additional donor vibronic levels up to the order $n_c$, where the overlap between the donor and acceptor vibronic states remains significant.   

Finally, we comment on how to increase the transfer rate for our state preparation scheme while preserving the fidelity of the target state. Since the gate time is controlled by the spin-spin coupling strength $J$, we also need to increase other system parameters, including $\omega_0$, $g$, $\Delta E$, $B$, and $\gamma$, for higher $J$. Although there are no strict upper limits on these parameters for the experimentally achievable range in a typical trapped-ion system of $J\lesssim2\pi\times1$ kHz, we still need to simultaneously maintain $\omega_0\sim g$ to satisfy the perturbation criterion. Hence, the limiting factor for our scheme is the strength of the spin-boson coupling $g$, which typically demands an order of magnitude higher in laser power for its generation compared to other parameters for the same strength when mapped to an analog trapped-ion quantum simulator. To mitigate this limitation, we can use analog Trotterization to simulate the extended ET Hamiltonian \cite{macdonell2021simulation,whitlow2023conical}. We can then maximize the laser power used to engineer the terms in the composite Hamiltonian by alternating between high power-consuming terms like $J_{ij}$ and $g$ or ET Hamiltonian and interaction Hamiltonian in the experimental sequence. With optimization of the sequence design, we essentially trade minimal non-commuting errors for increased interaction strengths.

\subsection{Generating $\ket{W_4^2}$ with Noises}

\begin{figure}[t]
    \includegraphics[width=1\linewidth]{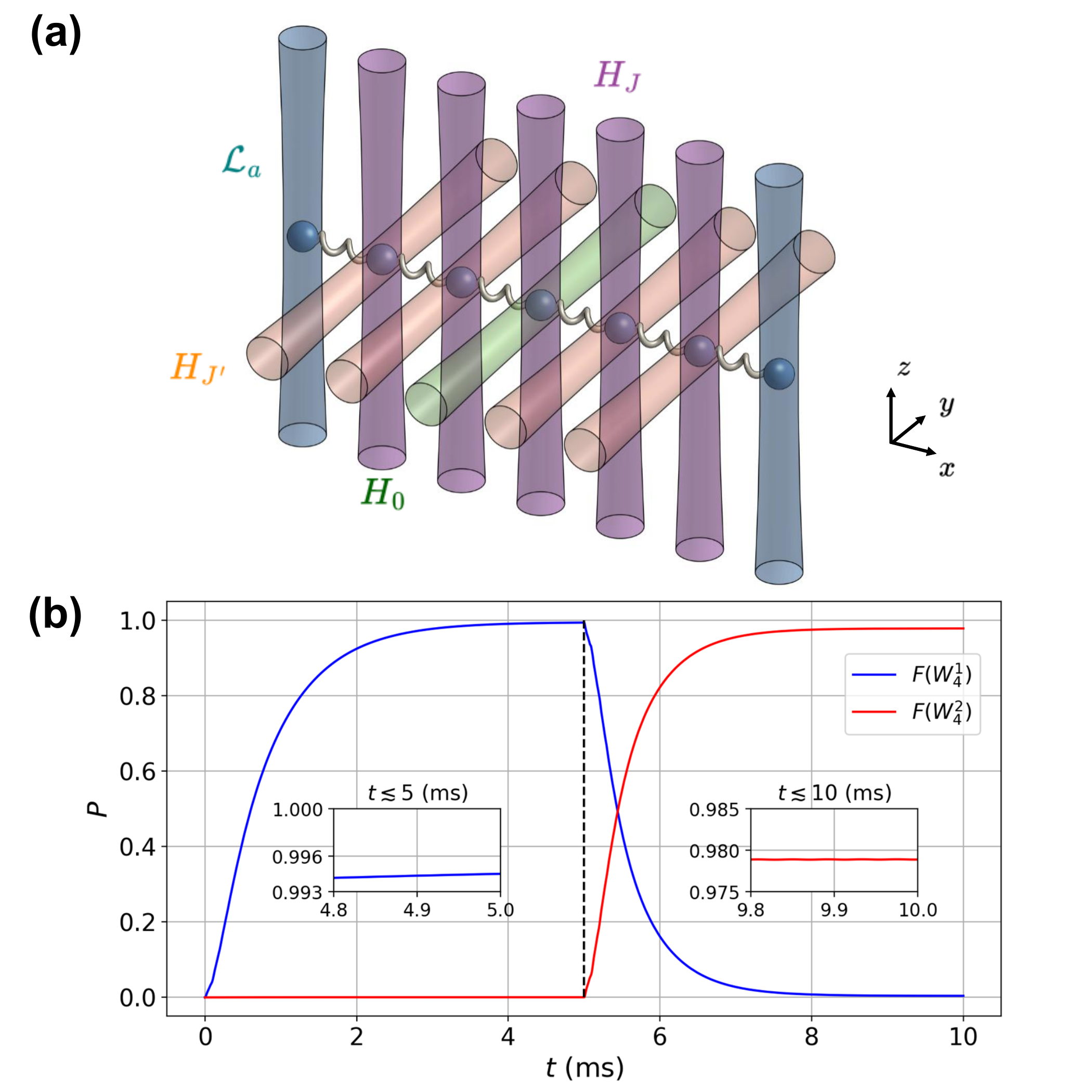}
    \caption{\justifying (a). Proposed experimental setup for implementing the protocol to engineer the four-qubit Dicke state with two excitations using a chain of seven ions, whose bosonic degrees of freedom are shared (represented by connecting springs). The center-most ion acts as the control qubit undergoing the non-interacting ET Hamiltonian $H_0$ induced by the green light fields, while the two edge ions are continuously cooled by the blue light fields to remove the bosonic excitations from the system. All five qubits interact with the purple light fields to create the effective spin-hopping Hamiltonian $H_J$ using the bosonic modes along the $z$ trap axis. However, only the target qubits are simultaneously addressed by the orange light fields, which generate the spin-spin interactions $H_{J'}$ among themselves using the $y$-direction bosonic modes. With appropriate laser parameters, the resultant couplings, given by $H_{\rm int} = H_J + H_{J'}$, are the required selective pairwise spin-hopping interactions. (b). State population dynamics of the protocol
    . The solid blue and red lines are the fidelities of the intermediate state $\ket{W_4^1}$ and the target state $\ket{W_4^2}$, respectively. The black dashed line marks the time at which the repumping $\pi$ pulse is implemented.    
     }\label{fig, W42_ex}
\end{figure}

To demonstrate the feasibility of our proposed setup, we numerically simulate the hybrid protocol for preparing $\ket{W_4^2}$ Dicke state using the parameters, given by a realistic experimental setup with a chain of seven $^{171}{\rm Yb}^+$ ions. In the experimental design, we use the five central ions as the qubits and the two edge ions as the coolants (see Fig. \ref{fig, W42_ex}(a)).
We shall identify the qubit ions from left to right with the indices $i=1$\textendash$5$. The ion at the center of the chain with the index $3$ is assigned as the control qubit, while the other qubit ions are referred to as the target qubits.
We also choose the trapping configuration so that the two radial mode spectra are well-separated into non-overlapping low-frequency and high-frequency radial mode sets. The realistic system parameters we consider are $(\Delta E,g) =(2,1)\omega_0$ with $\omega_0=2\pi\times10\text{ kHz}$. Suppose we generate the two-body coupling terms via the MS scheme using the Raman laser beams that propagate along the trap axis associated with the low-frequency radial mode set ($z$-direction in Fig. \ref{fig, W42_ex}(a)), a near-balanced coupling strength between the control qubit and target qubits of $J_{3,j}\equiv J$ can be achieved by the optimization of the individual laser beam powers and the global laser frequency detuning. Here, we consider the complete form of the resultant spin-hopping coupling $\sum_{i<j=1}^{5}J_{ij}\sigma_i^x\sigma_j^x+B\sum_{i=1}^5{\sigma_i^z}$ with $B=0.6\omega_0$, where the second term describing the transverse field is added to suppress the off-resonant, double excitation processes. Simultaneously, we propose to implement the same scheme on the target qubits using the Raman laser beams that propagate along the other trap axis associated with the high-frequency radial mode set ($y$-direction in Fig. \ref{fig, W42_ex}(a)) to generate the $J_{ij}'$ terms that minimize the undesired $J_{ij}$ terms with $i,j\neq3$. By combining the two spin-hopping interactions, we shall obtain the selective pairwise spin-hopping interaction with $J\sim0.04\omega_0$. We account for possible experimental imperfections in interaction engineering by including the coupling strength imbalances of the order of $\delta J\sim0.01J$ and the residual couplings among the target qubits of $J_{\text{res}}\sim 0.001 J$ to our simulation (see Appendix \ref{Ap_ss} for the details of these coupling parameters). The dissipation rate $\gamma$ is set to $2V_{\text{e}}^m\sim0.07\omega_0$, which leads to an approximately optimal transfer rate for each pumping step. We also consider the boson heating of $\bar{n}=0.05$, which is a typical value when the third highest frequency mode of the $y$-direction radial mode set is used for spin-boson coupling and cooling. This mode is the preferred choice because of its adequate couplings to the central and edge ions as well as its robustness to environmental noises in a realistic, experimental setting. The heating can be effectively suppressed by setting $\Delta E = 2\omega_0$, as discussed in Sec \ref{sec_EET}. A $\pi$ pulse with the gate time of $1\; \mu\text{s}$ is implemented to repump the control qubit back to the $\ket{D}$ state after the $\ket{W_4^1}$ state is prepared in the target qubits. 

To test the performance of the protocol, we numerically integrate the master equation in Eq. \eqref{eq_master} and plot the population dynamics of the intermediate state $\ket{W_4^1}$ and the target state $\ket{W_4^2}$ in Fig. \ref{fig, W42_ex}(b), where $\ket{W_4^1}$ is obtained in $5$ ms with $99.4\% $ fidelity, and $\ket{W_4^2}$ is obtained in $10$ ms with $97.9\% $ fidelity. A more significant decrease in fidelity arises in the second iteration because the second pumping step is more susceptible to off-resonant couplings, $\sigma_i^+\sigma_j^+ + \text{h.c.}$, compared to the first pumping step. Specifically, $\ket{W_4^1}$ is connected to a larger number of states through first-order transitions, unlike the initial state, where the target qubits are all in the $\ket{\downarrow}$ state. 

\section{Conclusions and Outlooks}\label{sec_con}

In this paper, we have designed an extended ET model, which employs a single-site ET system as a quantum control knob to manipulate the steady-state entanglement in a separate target quantum system. To gain a conceptual understanding of our scheme, we developed a reduced model with an open three-level quantum system, which can be solved analytically to predict the dynamics of the extended ET model in the perturbative limit. We showed that, with typical interactions available in trapped-ion systems, our protocol could be iteratively applied to a target system with $N$ qubits to generate Dicke spin states with an arbitrary number of excitations using either a hybrid or a fully dissipative scheme. A similar technique can also be utilized to generate a $W$ state in a target system of $N$ bosons. Furthermore, we presented the experimental setup for implementing the key interactions required for generating a Dicke spin state and analyzed the effects of the possible sources of noise that could arise in a realistic setting. Finally, we provided an example of an experimental setup for generating the $\ket{W_4^2}$ Dicke state using the hybrid scheme and numerically tested the performance of the scheme in the presence of typical decoherence sources. 


Unlike previous designs, our proposed framework offers the capability to dissipatively engineer $N$-qubit and $N$-boson $W$ states as well as Dicke $N$-qubit states, whose resilience to typical experimental imperfections is desired in quantum information science.
Although the structure of the target state is constrained by the types of Hamiltonians that can be engineered on a given experimental platform, the general protocol we have proposed can serve as a powerful tool for the deterministic preparation of entangled states. In particular,
the flexibility of our approach will allow extensions to the generation of entangled states in both spin and bosonic sectors of trapped-ion systems. These states can be used as resources for quantum squeezing and amplification, relevant to a wide range of metrology applications \cite{ma2011quantum,zhang2014quantum,PRXQuantum.3.010202}. Similarly, the ability to initialize different classes of entangled states on trapped-ion systems is a practical resource for simulating signature behaviors of complex many-body quantum models in high-energy physics, condensed matter, and biochemistry. For instance, a recent study has shown that entanglement can speed up the transfer dynamics of excitation from the donor monomer to the acceptor monomer, each composed of several molecular sites, in Frenkel-exciton systems. Specifically, the most efficient transfer occurs when the excitation of the donor monomer is delocalized in the $W$ state \cite{padilla2025vibrationallyassistedexcitontransfer}. 

Moreover, it is worth further studying the possibility of engineering other classes of entangled states in more complex structures using our proposed framework. For instance, carefully designed interactions between the control qubit and the target system may enable the generation of cluster states, which are particularly valuable for various quantum computing algorithms \cite{nielsen2006cluster,lanyon2013measurement}. The capabilities of our protocol can also extend beyond creating entanglement in qubits and bosons as there is, in principle, no restriction on the characteristics of the degrees of freedom contained in the target system. For example, our protocol can be generalized to qudits and used to create different classes of $W$ qudit states with qubit-qudit couplings, which are available in trapped-ion platforms \cite{edmunds2024constructingspin1haldanephase}. Along with the possibilities of modifying the control system beyond the single-site ET model and combining it with other state-preparation techniques, such as measurement-based protocols \cite{verresen2021efficiently,halder2021measurement,PhysRevX.14.021040,iqbal2024topological}, engineering jump operators of more complex forms \cite{Reiter2012,weimer2016tailored}, and implementing multiple independent dissipation channels \cite{wang2024topologically,jiao2025protecting,so2025quantumsimulationschargeexciton}, this framework opens up many opportunities for future investigations.

\vspace{0.5cm}

\begin{acknowledgments}
The authors acknowledge Yilun Xu for his contributions to the development of the reduced three-level formalism and for verifying related results. We also thank Diego Fallas Padilla for insightful discussions.

G.P. acknowledges the support of the Welch Foundation Award C-2154, the Office of Naval Research Young Investigator Program (Grant No. N00014-22-1-2282), the NSF CAREER Award (Grant No. PHY-2144910), the Army Research Office (W911NF22C0012), and the Office of Naval Research (Grants No. N00014-23-1-2665 and N00014-24-12593). We acknowledge that this material is based on work supported by the U.S Department of Energy, Office of Science, Office of Nuclear Physics under the Early Career Award No. DE-SC0023806. H.P. acknowledges support from the NSF (Grant No. PHY-2207283) and from the Welch Foundation (Grant No. C-1669).

\end{acknowledgments}
\bibliography{new_refs}{}
\begin{figure*}[t!]
    \includegraphics[width=1\linewidth]{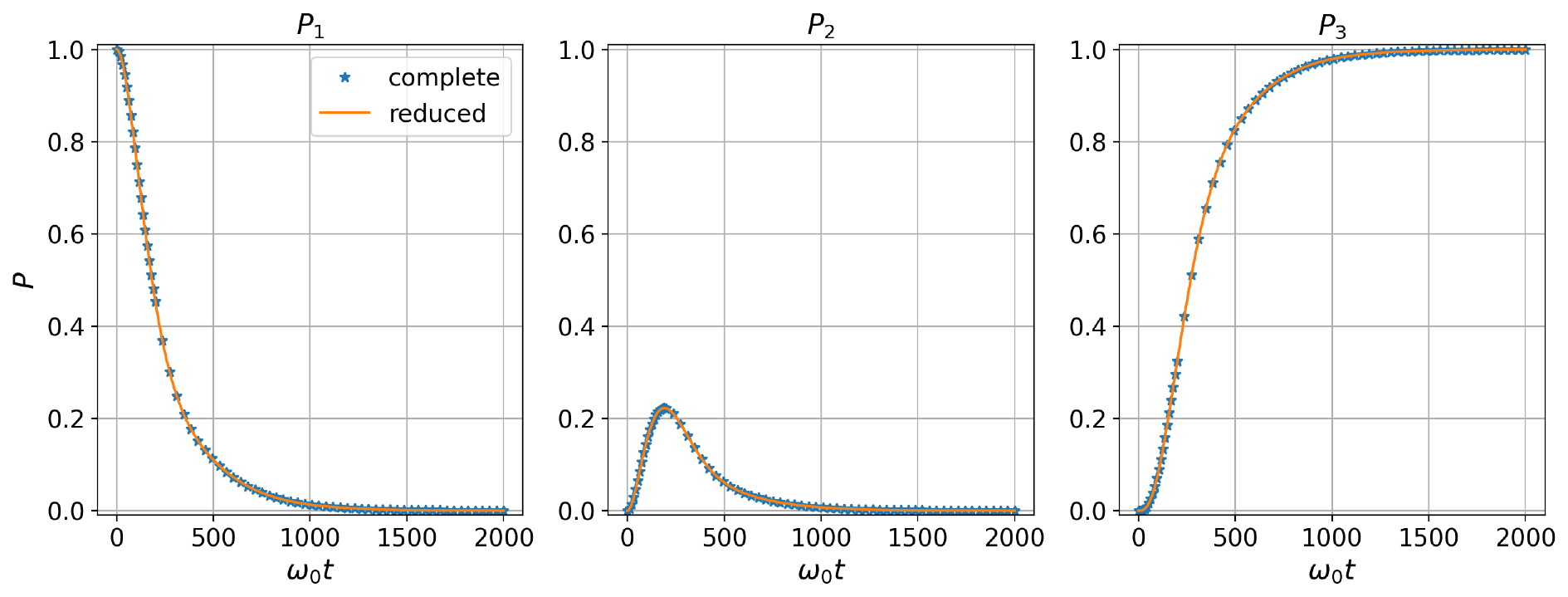}
    \caption{\justifying Population dynamics of states $\ket{1}$, $\ket{2}$, and $\ket{3}$ (from left to right) obtained by analytically solving Eq. \eqref{dif_3} (reduced model) and numerically integrating Eq. \eqref{eq_master} (complete model) with $\Delta E= \omega_0, V = 0.01 \omega_0, g=\omega_0,\gamma=|V_{\text{e}}|,$ boson cutoff at $n_c= 10$. We have assumed $\langle e_a^{j_1}|e_d^{i_1}\rangle=1$ to recover the original ET model. The results for the reduced model are plotted with solid lines, and those for the complete model are plotted with points. The system is initialized in the state $\ket{1}$ and approaches the steady state $\ket{3}$ at $t \rightarrow \infty$. A good agreement between the results obtained from the two models indicates the validity of the approximations. }
    \label{fig:compare}
\end{figure*}
\appendix
\section*{Appendix}

\subsection{Verification of Reduced Dynamics}\label{Ap_A}
\begin{figure*}[htbp]
    \includegraphics[width=0.9\linewidth]{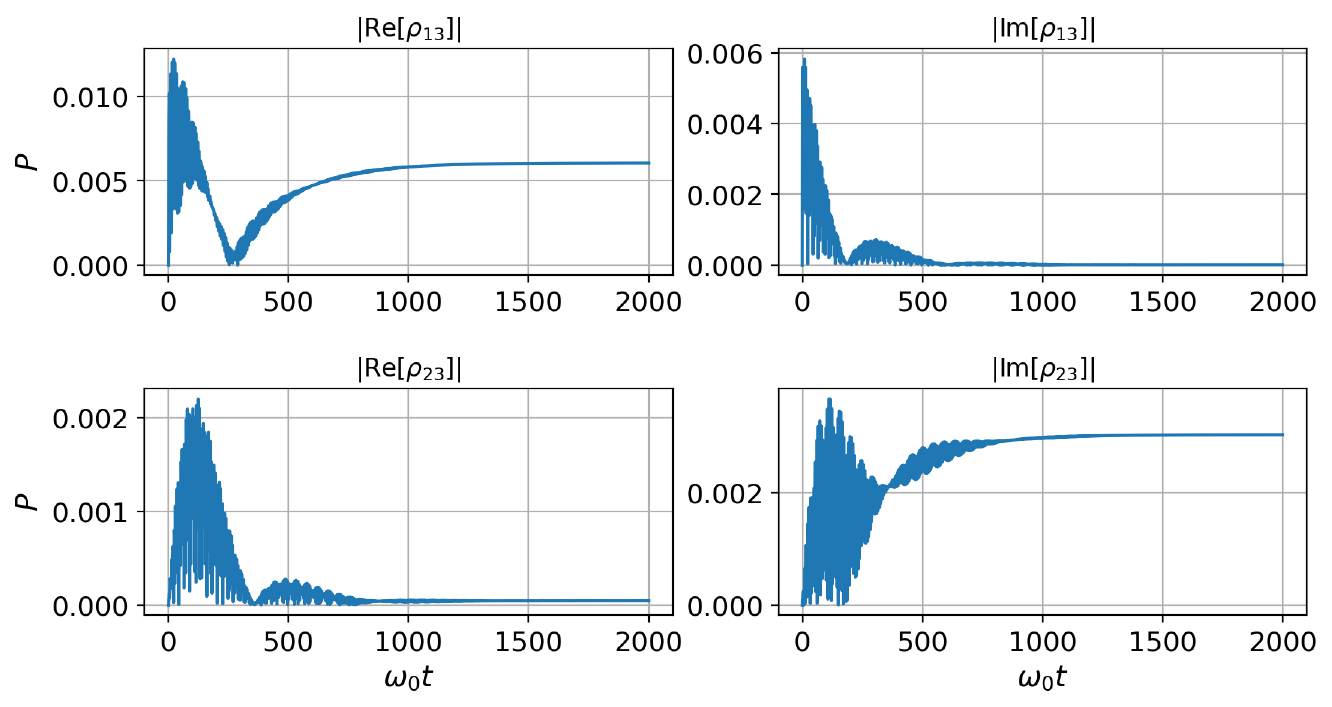}
    \caption{ Time evolution of coherences between states $\ket{1}$ and $\ket{3}$ and between states $\ket{2}$ and $\ket{3}$.}
    \label{fig:a_err}
\end{figure*}

To check the validity of Eq. \eqref{dif_3}, we consider the simple case of $\langle e_{i_2}|H_\text{int}|e_{i_1}\rangle=1$ such that the extended ET model in the truncated basis becomes equivalent to the single-site model. 
We numerically integrate Eq. \eqref{eq_master} with the complete ET Hamiltonian defined in Eq. \eqref{ET_H} for $H_{\text{ext}}=0$ and compare the population dynamics of states $\ket{1},\ket{2},$ and $\ket{3}$ with those obtained by analytically solving the differential equations in Eq. \eqref{dif_3}. The results plotted in Fig. \ref{fig:compare} show good consistency between the two methods in the perturbative regime. We also plotted the coherences $\rho_{13}$ and $\rho_{23}$ as a function of time in Fig. \ref{fig:a_err}, showing that they are not significant compared to the on-resonance dynamics. 
As long as the perturbation conditions in Eq. \eqref{pc} are satisfied, the reduced model remains applicable to scenarios with non-zero $H_{\text{ext}}$ and generic $\langle e_{i_2}|H_\text{int}|e_{i_1}\rangle$, provided the effective $V_{\text{e}}$ being accurately calculated.

\subsection{Increasing Fidelity at Finite Temperature}\label{ap_FT}
\begin{figure}[ht]
    \includegraphics[width=0.95\linewidth]{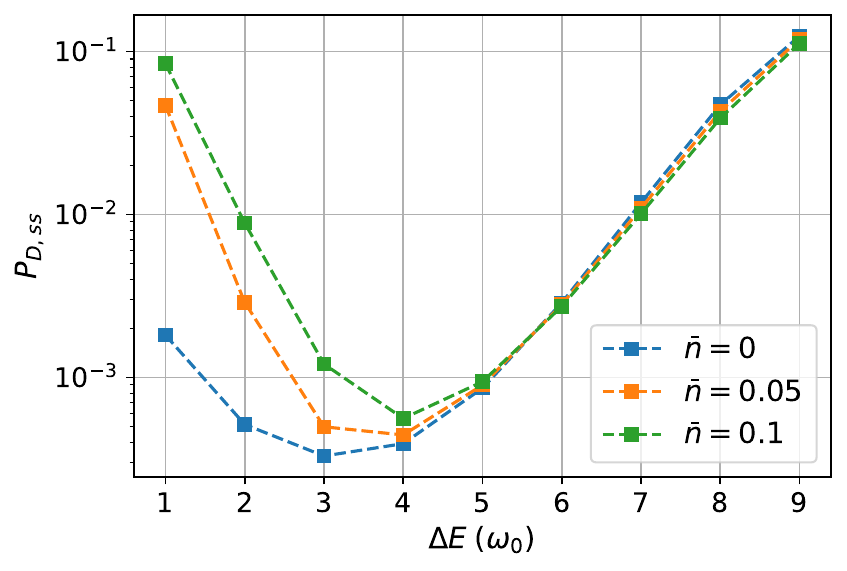}
    \caption{\justifying Steady-state donor population as a function of $\Delta E$. The data points are sampled under resonance condition $\Delta E = n\omega_0$. The population at different temperatures $ \bar{n}=0,0.05,0.1$ are plotted with blue, orange, and green colors, respectively. 
     }\label{fig:f_plot}
\end{figure}

In typical experimental conditions of the schemes proposed in the main text, the Markovian bath engineered via sympathetic cooling inevitably has a nonzero temperature (\(\bar{n} \neq 0\)), which reduces the target state fidelity. However, this effect can be mitigated by increasing \(\Delta E\), thereby enhancing the couplings between the initial donor vibronic state and acceptor vibronic states with sufficiently high excitations.

To elaborate on this point, we study the steady-state population of the acceptor state $\ket{A}$ in the single-site ET model, which is directly related to the fidelity of the target state in the proposed schemes. The dynamics at $\bar{n}=n_0\neq0$ can be qualitatively understood in a similar way as the for zero temperature case mentioned in Section \ref{sec_SET} of the main text. Let $n_e$ be the cutoff of the excitation number such that all acceptor vibronic states with vibrational quantum number $n>n_e$ excitations have negligible populations. Assuming the resonance condition $\Delta E= k\omega_0$, if $k>n_e$, all levels in the acceptor thermal state become off-resonant, and a nearly complete transfer of the donor population is possible again. On the other hand, if $k<n_e$, the presumed acceptor steady state will have $n_e-k$ levels energetically resonant with the donor eigenstates such that the transition back to the donor state is still allowed by the donor-acceptor coupling $V\sigma_x$. In this case, the system will stabilize to a mixture of donor and acceptor states, leading to a lower fidelity.

However, for a fixed spin-boson coupling \( g \), increasing \( \Delta E \) reduces the overlaps between donor and acceptor vibronic states, thereby decreasing the effective coupling strengths. This, in turn, increases the steady state donor population. Specifically, when the resonant couplings become sufficiently weak, two additional factors must be considered: \emph{(i)} off-resonant couplings between donor and acceptor vibronic states become comparable to the ``resonant'' coupling, and \emph{(ii)} the fraction of the population that remains in a given vibronic state upon the action of the jump operator \( a \) (or \( a^\dagger \)) is finite because of the nonzero diagonal elements of \( a \) and \( a^\dagger \) in the unperturbed vibronic basis in Eq. \eqref{lev_HET}. 
These two effects constitute the fundamental limitation to the fidelity even at zero temperature $\bar{n}=n=0$.

The interplay between off-resonant couplings and the specific form of the jump operators in the unperturbed basis tends to increase the population remaining in the donor state. When \( \Delta E \) becomes sufficiently large, these factors dominate over the resonant couplings, and we expect a reversal of the steady-state donor population. This suggests the existence of an optimal \( \Delta E \) that maximizes the acceptor state population.
      
We numerically verify the above arguments by solving the steady-state master equation, which is obtained by setting the LHS of \eqref{eq_master} to zero. We  plot the steady-state donor population $P_{D,ss}$ as a function of resonant $\Delta E =n\omega$ at different temperatures in Fig. \ref{fig:f_plot}. For all the three selected temperatures, before the optimal $\Delta E$ is reached, $P_{D,ss}$ decreases as $\Delta E$ increases, indicating an increase of the acceptor state fidelity. The improvement of fidelity is more significant in the cases of $\bar{n}\neq0$, demonstrating the effectiveness of this technique at finite temperatures. 

\subsection{Steady-state $W$ Boson State}\label{ap_eph}
\begin{figure}[h]
    \includegraphics[width=0.95\linewidth]{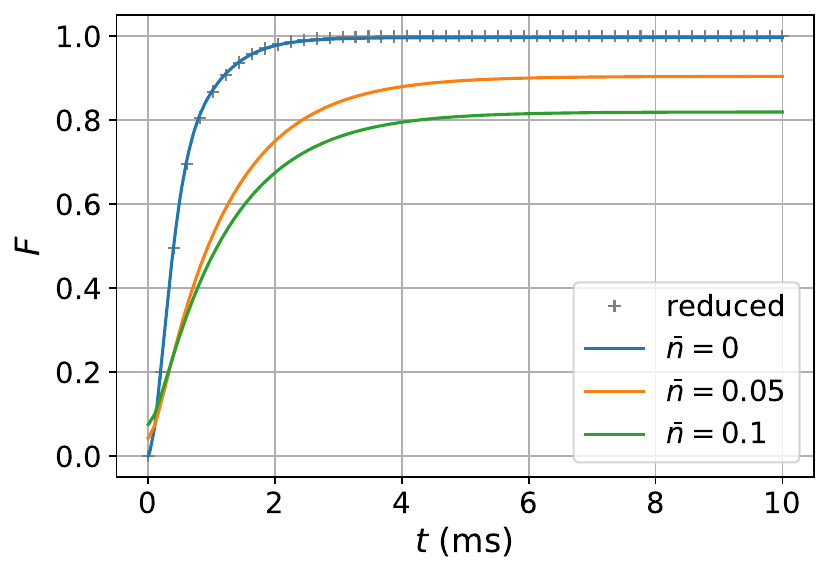}
    \caption{\justifying Target state fidelity as a function of time of the protocol for engineering a $\ket{W_2^1}$ boson state. The blue, orange, green solid lines plot the fidelities at bath temperatures $\bar{n}=0,0.05,0.1$. The reduced 3-level solution for $\bar{n}=0$ is plotted with the grey dots. 
     }\label{ep_plot}
\end{figure}

We demonstrate an example scheme of engineering the steady-state $N=2$ (triplet) $W$ boson state. We plot the fidelities of the target state at different boson temperatures $\bar{n} = 0, 0.05,0.1$ in Fig. \ref{ep_plot}.  The results are obtained by numerically integrating the complete model described by Eq. \eqref{eq_master} of the main text. Here $H$ is given by Eq. \eqref{H_eph} in the main text with parameters set to  $(g,V)=(1,0.05)\omega_0$ for $\omega_0=2\pi\times10\;\text{kHz}$. The spin energy splitting $\Delta E$ is set to $\omega_0$ for $\bar{n}=0$ and $2\omega_0$ for $\bar{n}=0.05,0.1$ to increase the fidelity of the target state in the latter cases. The dissipation rate is set to $\gamma=2V_{\text{e}}=2\sqrt{2}\tilde{g}e^{-\tilde{g}/2\omega_0}V\gtrsim0.05\omega_0$ to optimize the transition rates. 

Initially, the system is in a product state $\rho_0=\ket{D}\bra{D}\otimes (U\rho_{th}(n_0)U^\dagger)\otimes\rho_{th}(n_0)\otimes\rho_{th}(n_0)$, where $\rho_{th}(n_0)$ is the single-boson thermal state characterized by an average boson number $n_0$, and $U=\exp(\tilde{g}/2(a^\dagger-a))$ is the displacement operator. Again, we assume the damped bosonic mode is initially in thermal equilibrium with the bath ($n_0=\bar{n}$). The steady-state fidelities in the three cases are $99.6\%,90.4\%,81.9\%$, demonstrating that the scheme is sensitive to the boson temperature.   

\subsection{Experimental Spin-spin Coupling Strength}\label{Ap_ss}
The numerical values of the selective pairwise spin-hopping coupling strength $J_{ij}$ used for simulating the dynamics in Fig. \ref{fig, W42_ex} of the main text are listed in the following table (in units of kHz):
\begin{scriptsize}
\[
\begin{array}{|c|c|c|c|c|c|}
    \hline
    & j=1 & j=2 & j=3 & j=4 & j=5 \\ \hline
    i=1 & 0 & 2.25\times10^{-4} & 0.398 & 1.07\times10^{-4} & 1.15\times10^{-4} \\ \hline
    i=2 & 2.25\times10^{-4} & 0 & 0.402 & 1.03\times10^{-4} & 1.07\times10^{-4} \\ \hline
    i=3 & 0.398 & 0.402 & 0 & 0.402 & 0.398 \\ \hline
    i=4 & 1.07\times10^{-4} & 1.03\times10^{-4} & 0.402 & 0 & 2.25\times10^{-4} \\ \hline
    i=5 & 1.15\times10^{-4} & 1.07\times10^{-4} & 0.398 & 2.25\times10^{-4} & 0 \\ \hline
\end{array}
\]
\end{scriptsize}
\noindent This interaction matrix is generated by taking the difference between the spin-spin coupling matrix acting on all the qubit ions and one that applies only to the target qubits, both calculated using Eq. (24) in \cite{monroe2021programmable}. As described in the main text, the opposite signs of the two matrices come from the imposed conditions of the motional phases. Prior to taking the difference, each matrix is first optimized through the individual ion-laser Rabi coupling strength $\Omega_i$ such that the off-diagonal matrix elements are balanced and minimally close to the desired value (in this case, 0.400 kHz).

\subsection{Dissipative Generation of GHZ Spin States}\label{Ap_3}

Consider a target system with $2N$ spin-$1/2$ qubits labeled by indices $i = 1, 2, \ldots, 2N$. Consider an interaction in the form $V\sigma_0^x+H_{\text{ex},z}\sigma_0^z$. In this case, it is more convenient to  absorb the second term into the unperturbed Hamiltonian such that 
\begin{equation}
    H_0^\prime=\frac{\Delta E}{2}\sigma_0^z+H_{\text{ex},z}\sigma_0^z+H_{\text{ext}}+\frac{g}{2}\sigma_0^z\left(a+a^\dag\right)+\omega_0a^\dag a,
\end{equation}
and the simple interaction $V\sigma_0^x$ acts as a perturbation to the system.
The new vibronic donor/acceptor states, which are the unperturbed eigenstates of $H_0'$, can then be written as:
\begin{equation}
\left|D,n_d,i\right\rangle=\left|D,n_d\right\rangle\otimes\left|e_d^i\right\rangle,\ \ \left|A,{n}_a,j\right\rangle=\left|A,n_a\right\rangle\otimes\left|e_a^j\right\rangle,
\end{equation}
 where $\{\ket{e_d^i}\},\{\ket{e_a^j}\}$, labeled by quantum numbers $i$ and $j$, are the sets of eigenstates of the two effective external Hamiltonians $H_{\text{ext}}^{+}$ and $H_{\text{ext}}^{-}$:
 \begin{equation}
H_{\text{ext}}^{\pm}=H_{\text{ext}}\pm H_{{\rm ex},z}. \label{H_ex}
 \end{equation} 
They correspond to the cases in which the control qubit is in either $\ket{D}$ or $\ket{A}$ state. 

We aim to find some $H_{\text{ex},z}$ that shifts the spectrum of $H_{\text{ext}}$ depending on the state of the control qubit. This shift should ensure that one of $H_{\text{ext}}^+$ or $H_{\text{ext}}^-$ has a set of product states that are easy to prepare as its eigenstates, while the other one has an eigenstate with the desired entanglement. One possible choice is to generate the following interactions: 
\begin{eqnarray}
    H_{\text{ex},z} &=& \frac{E_0}{2}\sum \nolimits_{i=1}^{2N}\sigma^z_i, \nonumber\\
    H_{\text{ext}} &=& \frac{E_0}{2}\sum \nolimits_{i=1}^{2N} \sigma^z_i +
    \frac{k}{2}\left(\prod \nolimits_{i=1}^N\sigma^x_{i} +\prod \nolimits_{i=N+1}^{2N}\sigma^x_{i} \right),\nonumber\\
\end{eqnarray}
where $\sigma_i^z,\sigma_i^x$ are Pauli operators acting on the $i$-th spin. Under the limit of $E_0\gg k$, the highest-energy eigenstate of $H_{\text{ext}}^+$, to the first order, is approximately $\ket{e_d^{max}} = \ket{\uparrow\ldots\uparrow}_{2N}$ with $E=2NE_0$. On the other hand, $H_{\text{ext}}^-$ has an eigenstate $\ket{e_a^{s}}=\frac{1}{\sqrt{2}}(\ket{\uparrow\ldots\uparrow}_{2N}-\ket{\downarrow\ldots\downarrow}_{2N})$ with zero energy. If the system is initialized in a donor vibronic state 
$\ket{\psi_i}=\ket{1}\equiv\ket{D,n_d=0}\otimes \ket{e_d^{max}}$ and $\Delta E = \omega_0-2NE_0 $, the only resonant state that has non-zero overlap with $\ket{\psi_i}$ in the perturbative regime, where $\omega_0\gg NE_0$ and $k\gg V$, is the acceptor vibronic state $\ket{2}\equiv\ket{A,n_a=1}\otimes \ket{e_a^{s}}$. Under boson dissipation, the system will then be dissipatively driven to $\ket{3} \equiv\ket{A,n_a=0}\otimes\ket{e_a^s}$. The master equation dynamics can be described by the formalism in Sec. \ref{sec_EET} of the main text with effective Rabi frequency $V_{\text{e}}=-V \tilde{g} e^{(-\tilde{g}^2/2)}/\sqrt{2}$. Through spin-spin interaction $\sigma_0^z\sigma_i^z$ and $N$-spin interaction $\prod_i^N \sigma_i^x$, we can prepare a $2N$-spin GHZ state in the steady state starting from a simple product state. Implementing the spin-spin MS interaction enables the generation of two-spin and four-spin GHZ entangled states using our protocol. We shall discuss the former case in details since the four-spin case involves implementing $\sigma^i_z\sigma^j_z$ and $\sigma^i_x\sigma^j_x$ simultaneously.
\begin{figure*}[ht]
    \centering
    \includegraphics[width=0.8\linewidth]{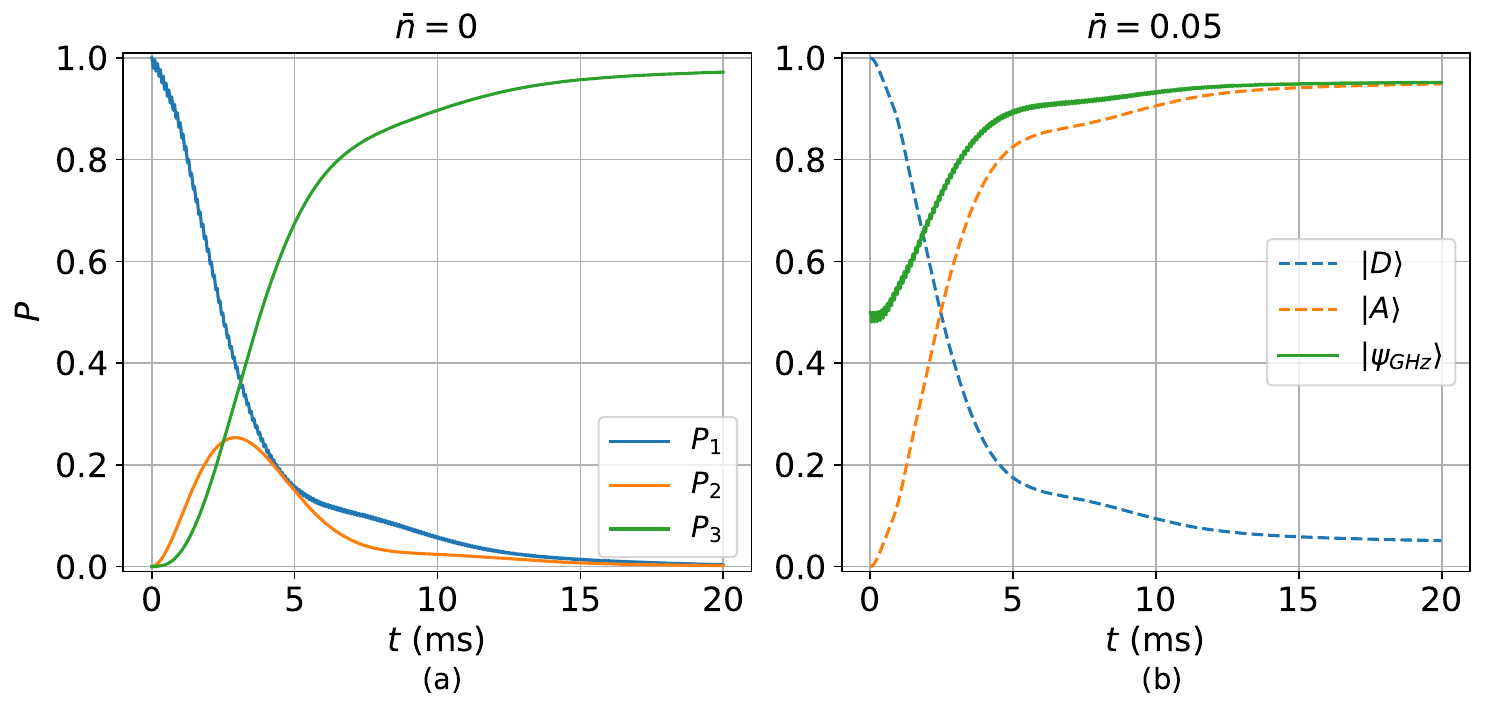}
    \caption{\justifying State population dynamics of the protocol for engineering a two-spin GHZ state. (a) Population of states $\ket{1},\ket{2},$ and $\ket{3}$ as a function of time without boson heating ($\bar{n}=0$). (b) Spin population and target state fidelity as a function of time with boson heating ($\bar{n}=0.05$). The dashed lines are the populations of the states $\ket{\uparrow}$ and $\ket{\downarrow}$, respectively. The solid line is the fidelity of the target state $\ket{\psi_\text{GHZ}}=\frac{1}{\sqrt{2}}(\ket{\uparrow\uparrow}-\ket{\downarrow\downarrow})$.
     }\label{ghz_plot}
\end{figure*}

For the two spin case, the total Hamiltonian takes the form
\begin{eqnarray}
     H_{\rm tot}&=&\frac{\Delta E}{2}\sigma_0^z+V\sigma_0^x+\frac{E_0}{2}\sigma_0^z(\sigma_1^z+\sigma_2^z)+\frac{E_0}{2}(\sigma_1^z+\sigma_2^z)\nonumber\\
    &+&\frac{k}{2}(\sigma_1^x+\sigma_2^x)
    +\frac{g}{2}\sigma_0^z\left(a+a^\dag\right)+\omega_0a^\dag a.
    \label{H_ghz}
\end{eqnarray}
We test the effectiveness of this protocol in the ideal case without boson heating ($\bar{n}=0$) and a more realistic case of $\bar{n}=0.05$ by numerically integrating Eq. \eqref{eq_master} with $H_{\rm tot}$ defined in Eq. \eqref{H_ghz}. We choose a set of experimentally accessible parameters: $(\Delta E,E_0,k,g,V)=(1.4,0.2,0.04,0.5,0.008)\omega_0, \gamma = 2V_{\text{e}}, \omega_0 = 2\pi\times25 \;\text{kHz} $. The state dynamics are plotted in Fig. \ref{ghz_plot}. The target state $\ket{\psi_\text{GHZ}}=\frac{1}{\sqrt{2}}(\ket{\uparrow\uparrow}-\ket{\downarrow\downarrow})$ is obtained in $20\;\text{ms}$ with $97.2\%$ fidelity without heating and  $95.1\%$ fidelity with heating.  

The fidelity of this protocol depends on the limit of $E_0\gg k$ since the initial product state will no longer be an approximate eigenstate of $H_{\text{ext}}^d$ otherwise. Additionally, the perturbation condition requires $k\gg V$. However, $V$ controls the transfer rate of the protocol, which makes a large $V$ desirable. Thus, the optimal gate time will be constrained in two ways. First, the typical two-body coupling coefficient $E_0$ achieved in experiments is $\sim 2\pi\times1\;\text{kHz}$, which is small. Second, though $\omega_0$ can be tuned to a large value because it is controlled by the laser frequency as long as $\omega_0\sim g$, the spin-boson coupling $g$ is limited by the laser power. Moreover, we want $\tilde{g}=g/\omega_0\sim 1$ to maintain a large overlap between the donor and acceptor states. Therefore, the choice of $V$ is also limited by the maximum $g$ that can be implemented, which is usually $\sim2\pi\times10\;\text{kHz}$. 

\end{document}